\newcommand{\Msun}{~M_\odot}
\newcommand{\Rsun}{~R_\odot}
\newcommand{\msun}{M_\odot}

\newcommand{\gcm}{\rm ~g~cm^{-3}}
\newcommand{\kms}{\rm ~km~s^{-1}}
\newcommand{\ergs}{\rm ~erg~s^{-1}}

\newcommand{\ml}{~\Msun ~\rm yr^{-1}}

\documentclass[11pt,preprint]{aastex}

\begin{document}

\title{YOUNG CORE COLLAPSE SUPERNOVA REMNANTS AND THEIR SUPERNOVAE}
\author{Roger A. Chevalier}
\affil{Department of Astronomy, University of Virginia, P.O. Box 3818, \\
Charlottesville, VA 22903; rac5x@virginia.edu}

\begin{abstract}
Massive star supernovae can be divided into four categories
depending on the amount of mass loss from the progenitor star
and the star's radius: red supergiant
stars with most of the H envelope intact (SN IIP), stars with
some H but most lost (IIL, IIb), stars with all H lost
(Ib, Ic), and blue supergiant stars with a massive H envelope
(SN 1987A-like).
Various aspects of the immediate aftermath
of the supernova are expected to develop in different ways
depending on the supernova category:
mixing in the supernova, fallback on the central compact object,
expansion of any pulsar wind nebula, interaction with
circumstellar matter, and photoionization by shock breakout
radiation.  The observed properties of young supernova
remnants allow many of them to be placed in one of the
supernova categories; all the  categories
are represented except for the SN 1987A-like type.  
Of the remnants with central pulsars, the pulsar
properties do not appear to be related to the
supernova category.
There is no evidence that the supernova categories form a
mass sequence, as would be expected
in a single star scenario for the evolution.
Models for young pulsar wind nebulae expanding into supernova
ejecta indicate initial pulsar periods of $10-100$ ms and
approximate equipartition between particle and magnetic energies.
Ages are obtained for pulsar nebulae, including an age of $2400\pm 500$
yr for 3C58, which is not consistent with an origin in SN 1181.
There is no evidence that mass fallback plays a role in neutron
star properties.

\end{abstract}

\keywords{stars: neutron --- stars: mass loss --- supernovae: general --- 
supernova remnants}

\section{INTRODUCTION}

The discovery of the pulsar in the Crab Nebula showed that
neutron stars are born in core collapse supernovae and that
the PWN (pulsar wind nebula) is capable of sweeping up the immediately
surrounding supernova ejecta.
There has recently been a substantial increase in the number of
observed young supernova remnants containing pulsars.
In some cases, a pulsar was strongly suspected to be present, but
was only discovered because of the increased sensitivity to pulsars
afforded by new observatories such as the {\it ASCA} and 
{\it Chandra} X-ray
observatories.
These discoveries include PSR J0205+64 in 3C58 (Murray et al. 2002),
PSR J1846--03 in Kes 75 (Gotthelf et al. 2000), PSR J1811--19
in G11.2--0.3 (Torii et al. 1997), PSR J1124--59 in G292.0+1.8
(Camilo et al. 2002a), and PSR J1930+19 in G54.1+0.3 (Camilo et al. 2002b).
In addition to the PWN, the interaction of the supernova with
its surroundings has been observed in some cases, showing a variety
of types of interaction.
These combined observations give the possibility of relating the
supernova remnant to the observed supernova types.
It might also be expected that the properties of both the central
neutron star and the surrounding PWN would depend on the type
of supernova responsible for their formation.

In addition to the pulsar discoveries, there is also a growing class
of compact central objects in supernova remnants that are not
normal pulsars.
These are likely to be neutron stars, and include radio quiet objects
like the source in Cas A (Pavlov et al. 2000) and the AXPs (anomalous
X-ray pulsars) such a the 12 s X-ray pulsar in Kes 73 (Vasisht
\& Gotthelf 1997).
In cases where these occur in a young remnant (age $\sim 1000$ yr),
there is the opportunity to investigate the supernova type.

Extensive supernova observations have clarified the various kinds
of core collapse supernovae.
Type IIP supernovae have a plateau light curve implying a massive H envelope,
Type IIL have a linear light curve, Type IIn have narrow emission lines,
SN 1987A-like have H in their spectra but are relatively faint and are powered
by radioactivity at maximum, and Type Ib/c are lacking H and are also
powered by radioactivity.
The Type IIb supernovae, which are SNe II (Type II supernovae)
that have the spectroscopic
appearance of a SN Ib/c at late times appear to be undergoing strong mass
loss and
 have a low mass H envelope at the time
of the explosion.
This is also true for the SNe IIL, which probably have more H envelope
mass than a SN IIb, but an envelope mass that is less than 
the mass in the  core of the progenitor.
Type IIn supernovae typically also have strong mass loss, but the
presence of narrow line emission may just depend on the sensitivity
to detect lines (e.g., narrow lines were observed from SN 1987A).
We thus consider 4 basic categories  of core collapse supernovae:
SN IIP, SN 1987A-like, SN IIL/b,  and SN Ib/c.
The first two categories  end their lives with massive H envelopes and
the next two have increasing amounts of mass loss.

Current information on supernova rates (Cappellaro et al. 1993, 1997)
does not strongly constrain the relative rates of these events, although
Cappellaro et al. (1997) find that SN 1987A-like events are not large
contributors to the rate of core collapse supernovae.  
The {\it observed} rate of SNe IIL is comparable to that of SNe IIP
(Cappellaro et al. 1993), but this might not reflect
the intrinsic rate.   From 
the theoretical point of view, Heger et al. (2003) have summarized
results on the end point of massive single stars.
In this view, SNe IIP supernovae come from stars of mass $\sim 9-25\Msun$, SNe Ib/c
from mass $\ga 35\Msun$, and SNe IIL/b from the intermediate range.
The rate of core collapse supernovae is strongly dominated by Type IIP
events; the ratio of IIP to IIL/b rates is $\sim 10$ and IIP to Ib/c $\sim 5$.
In addition, almost all of the SNe IIL/b and a substantial number
of SNe Ib/c undergo core fallback to a black hole and do not leave
a neutron star.
These relative numbers are changed if binaries make an important contribution
to Type IIL/b and Ib/c supernovae, which is a good possibility
(Nomoto et al. 1995; Wellstein \& Langer 1999).

There is some observational information on the masses of stars that give rise
to the various supernova types.
Images of the sites of SNe IIP 
have led to upper limits on the mass of
the progenitor of $15\Msun$ in 2 cases and $12\Msun$ in another
(Smartt et al. 2003).
A likely detection of the progenitor of the Type IIP SN 2003gd leads to
a zero age main sequence progenitor mass of $\sim 8-9\Msun$
(Van Dyk, Li, \& Filippenko 2003).
In the case of the Type IIb SN 1993J, the likely progenitor had an initial
mass of $13-22\Msun$ (Van Dyk et al. 2002).
The data on SNe IIP are consistent with the mass range expected for
single stars.
The mass of the SN 1993J progenitor is  lower than expected in the
single star scenario, but the likely binary companion has recently been
detected (Maund et al. 2004).
If SNe IIL/b and Ib/c came from single stars, more massive progenitors
than for SNe IIP would be expected.  From
the observed fact that a higher fraction of SNe Ib/c occur in Sc
galaxies than do SNe II, Cappellaro, Barbon, \& Turatto
(2004) infer a higher typical
initial mass for the SNe Ib/c than for the SNe II.
However, there may be considerable overlap in the initial masses for
these events.

In addition to variations in the progenitor mass, there are variations
in the supernova energy.
Although most core collapse supernovae are inferred to have an explosion
energy $\sim 1\times 10^{51}$ ergs, there has recently been found a class
of SNe with energies of a few $10^{49}$ ergs (Pastorello et al. 2004),
as well as some SNe Ib/c and SNe IIn with energies up to a few
$10^{52}$ ergs (e.g., Nomoto et al. 2001).
The rate of the low luminosity and low energy supernovae could be up to
$4-5$\% of the SN II rate (Pastorello et al. 2004).

The supernova type is important for determining the central conditions
in a supernova.
The compact core at the center of a
core-collapse supernova can be affected by
its immediate surroundings.
An aim of the present paper is to examine whether the central
evolution in the different kinds of core collapse supernovae and
the different aspects of circumstellar interaction can lead
to observable properties in the supernova remnant stage.
In \S~2, the density distribution, composition distribution, central
mass fallback,
circumstellar medium, and ionization at shock
breakout for the various supernova types are examined.
These properties can be important for the central compact object and
its surrounding supernova remnant.
The pulsar wind nebula interaction with surrounding supernova ejecta
is treated in \S~3.
The shock wave interaction with a red supergiant wind is in \S~4.
These models are applied to young supernova remnants with pulsar
wind nebulae in \S~5 and to remnants without pulsar nebulae in \S~6.
Where possible, the remnant properties are used to identify the
object with a supernova type.
The conclusions are  in \S~7.

\section{CORE COLLAPSE SUPERNOVA PROPERTIES}

\subsection{Density Distribution}

Both the inner PWN interaction and the outer circumstellar interaction
depend on the density structure of the supernova when it is in the
free expansion phase, with a velocity profile $v=r/t$.
This expansion phase should be reached within tens of days of the
explosion, and the structure is expected to depend on the supernova
type.

SN 1987A provides a well-studied case of an explosion of
a BSG (blue supergiant) and is the prototype of the SN 1987A-like class.
Models show that the central density gradient is small, so the density
can be expressed as $\rho_c=At^{-3}$, with $A\approx 10^9$
g cm$^{-3}$ s$^3$ (Woosley 1988; Shigeyama \& Nomoto 1990).
This result is suitable for a star with mass $\sim 18\Msun$ in which
the envelope mass is greater than the core mass, although there
has been some mass loss.
Another density estimate can be obtained from the analysis of
Matzner \& McKee (1999).
For a star with a radiative envelope, the explosion leads to
a density distribution with $\rho\propto r^{-1.06}$ at small radii.
The typical velocity of interest for a PWN is about $1000\kms$,
so the central density can be expressed as
\begin{equation}
\rho t^3=4.3\times 10^8\left(v\over 1000\kms\right)^{-1.06}
\left(M_{ej}\over 10\Msun\right)^{1.97}
E_{51}^{-0.97} {\rm~g~cm^{-3}~s^3},
\label{cden}
\end{equation}
where $M_{ej}$ is the ejected mass and $E_{51}$ is the explosion
energy in units of $10^{51}$ ergs.
For SN 1987A, the result is similar to that given above.
This expression is based on a harmonic mean analysis and misses
some of the inner density structure (Fig. 10 of Matzner \& McKee 1999),
but this part of the supernova is likely to be affected by
Rayleigh-Taylor instabilities in any case
(Kifonidis et al. 2003 and references therein).

SNe IIP are  expected to explode
as RSGs (red supergiants) with their
H envelopes nearly intact, giving rise to the extended plateau emission.
Matzner \& McKee (1999) give results for the
1-dimensional explosion of a $15\Msun$
star with its H envelope.
In the free expansion phase, there is a central region with
approximately constant density:
\begin{equation}
\rho t^3\approx 2.4\times 10^9 \left(M_{ej}\over 15\Msun\right)^{5/2}
E_{51}^{-3/2},
\end{equation}
where the scalings with ejected mass $M_{ej}$ and explosion
energy $E$ apply to stars with a similar initial density profile.

SNe IIL/b 
end their lives as red supergiants, but with small mass H envelopes
because of mass loss.
In this case, the stellar core gas is not effectively decelerated
by the envelope, so the the inner density structure should approximate
that of an exploded radiative core.
Model 3H11 of Iwamoto et al. (1997) for the Type IIb SN 1993J
has $M_{ej}=2.06\Msun$, of which $1.95\Msun$ is in the core, and
$E=10^{51}$ ergs, so that eq. (\ref{cden}) implies a density
of $1.7\times 10^{-5}$ g cm$^{-3}$ at $t=10^4$ s.
The model at this time shows a central density of $\sim 10^{-5}$ g cm$^{-3}$
(Fig. 2 of Iwamoto et al. 1997), in approximate agreement with eq. (1).
The expression for the explosion of a star with a radiative
envelope can also be applied to SNe Ib/c.
At the low mass extreme, the result for a star like the Type Ic
SN 1994I with $M_{ej}\approx 0.9\Msun$ (Iwamoto et al. 1994),
the coefficient in eq. (\ref{cden}) is $3.8\times 10^6$.
This shows the likely range in the central density of a supernova.

The density estimates given here apply to the inner, relatively
flat part of the supernova density profile.
At higher velocities, the density gradient steepens, eventually
going to a steep power-law profile (Matzner \& McKee 1999).
The profile begins to steepen at a characteristic velocity
\begin{equation}
v_b\approx 2\left(E\over M_{ej}\right)^{1/2}
=4500E_{51}^{1/2}
\left(M_{ej}\over 10\Msun\right)^{-1/2} \kms.
\end{equation}
At lower velocities, the density distribution is likely to be
relatively flat, but have some clumpiness as a result of
Rayleigh-Taylor instabilities.
A more specific result for the transition velocity comes from
assuming 2 power law segments ($\rho\propto r^{-m}$ on the inside and
$\rho\propto r^{-b}$ on the outside) with a sharp break between them.
The transition velocity is then (Chevalier \& Fransson 1992)
\begin{eqnarray}
v_{tr} & = & \left[{2(5-m)(b-5)\over (3-m)(b-3)}{E\over M_{ej}}\right]^{1/2}
\nonumber \\
& = & 3160\left[{(5-m)(b-5)\over (3-m)(b-3)}\right]^{1/2}
E_{51}^{1/2}
\left(M_{ej}\over 10\Msun\right)^{-1/2} \kms.
\label{vtr}
\end{eqnarray}
Matzner \& McKee (1999) find that the outer density profile can approximated
by a steep power law, with $b=10.2$ for a radiative star and
$n=11.7$ for a red supergiant.
For a star with $m=1.06$ and $b=10.2$, the reference velocity
becomes $3830\kms$.

An additional effect on the inner density distribution is that
radioactive $^{56}$Ni is present.
Deposition of the radioactive power over a period of days and weeks
results in the Ni bubble effect in which the radioactive material
expands and sweeps up nonradioactive gas.
The mass of $^{56}$Ni present in an explosion, $M_{Ni}$, can be estimated
from the late light curve.
The case of SN 1987A is especially well-determined, with
$M_{Ni}=0.075\Msun$.
Hamuy (2003) has estimated $M_{Ni}$ in the range $0.002-0.26\Msun$ for 20
Type IIP supernovae; he found that the objects with higher $M_{Ni}$
had higher explosion energies.
Results for 7 Type Ib and Ic supernovae indicate $M_{Ni}=0.07-0.15\Msun$
except for the highly energetic SN 1998bw with $M_{Ni}\approx 0.5\Msun$
(summarized by Hamuy 2003).
We take $M_{Ni}=0.1\Msun$ as a reference value and assume that $^{56}$Ni
is centrally located in region with a constant density distribution.
Before the Ni bubble effect, the outer velocity of the region is thus
\begin{equation}
V_0=363\left(M_{Ni}\over 0.1\Msun\right)^{1/3}
\left(\rho_a t^3\over 10^9 {\rm g~s^3~cm^{-3}}\right)^{-1/3}\kms,
\end{equation}
where $\rho_a$ is the density in the region.

The input of radioactive energy is $Q=3.69\times 10^{16}$ ergs g$^{-1}$
for $^{56}$Ni$\rightarrow ^{56}$Co with decay time
$\tau_d=8.8$ days and an additional
$7.87\times 10^{16}$ ergs g$^{-1}$
for $^{56}$Co$\rightarrow ^{56}$Fe with decay time
$\tau_d=110$ days.
Radiative diffusion is likely to be important for most of the
$^{56}$Co decays, so we assume that the energy input can be approximated
by that from $^{56}$Ni decays.
The outcome of the energy input is that the density of the Ni declines
and sweeps up a shell with final velocity $V_1$.
Conservation of energy shows that (Basko 1994)
\begin{equation}
\left(V_1\over V_0\right)^5-\left(V_1\over V_0\right)^2=
{5Q\over V_0^2}.
\end{equation}
A substantial Ni bubble effect requires that $5Q/V_0\gg 1$ or
\begin{equation}
140\left(M_{Ni}\over 0.1\Msun\right)^{-2/3}
\left(\rho_a t^3\over 10^9 {\rm g~s^3~cm^{-3}}\right)^{2/3}\gg 1.
\end{equation}
It appears that the effect is generally significant, except for
a low mass Type Ic supernova.
The final velocity of the bubble region is
\begin{equation}
V_1=975\left(M_{Ni}\over 0.1\Msun\right)^{1/5}
\left(\rho_a t^3\over 10^9 {\rm g~s^3~cm^{-3}}\right)^{-1/5}\kms,
\end{equation}
and the density contrast between the matter inside the bubble with
density $\rho_b$ and the ambient gas is
\begin{equation}
{\rho_b t^3\over\rho_a t^3}=0.052\left(M_{Ni}\over 0.1\Msun\right)^{2/5}
\left(\rho_a t^3\over 10^9 {\rm g~s^3~cm^{-3}}\right)^{-2/5}.
\end{equation}
The result can  be an order of magnitude reduction of the
central density.
The assumptions that the Ni is centrally located and sweeps all
the material into a shell lead to an underestimate of the actual
Ni bubble effect.
The shell is subject To Rayleigh-Taylor instabilities, but they
probably have a small effect on the shell expansion (Basko 1994).
More important is the fact that in a supernova like SN 1987A, the
Ni is mixed out from the center during the explosion, so the result
is a larger bubble with clumps of nonradioactive material inside
the bubble.

\subsection{Composition Distribution}

The composition distribution of SN 1987A has been extensively
investigated both theoretically and observationally.
A crucial aspect of mixing is the Rayleigh-Taylor instabilities
that occur when a denser core layer is decelerated by a lower density
outer layer (Kifonidis et al. 2003 and references therein).
The work of Kifonidis et al. (2003) included the density perturbations
expected in a supernova driven by the neutrino mechanism, and they found that
the instability in the Si/O interface is stronger than previous estimates.
However, the instability at the He/H interface is weak because of the
lack of significant perturbations in this region other than acoustic noise;
the result is that little H is mixed into the core region.
The observed late time line profiles of H$\alpha$ in SN 1987A show
a centrally peaked line, implying that H is present at low velocity
and that mixing to the center has occurred.
In modeling the emission,
Kozma \& Fransson (1998) find that H has been mixed down to velocities
$\la 700\kms$.
The inward mixing of H in SN 1987A is a puzzle as well as the outward
mixing of Fe.

In models appropriate for SNe IIP,
the massive H envelope is able to
decelerate the core material and the reverse shock wave carries H
back toward the center.
Shigeyama et al. (1996) note that that the He/C+O boundary is also
unstable, but that the instability at the He/H boundary
is much stronger and can mix in H to a velocity $\sim 1,000\kms$.
However, the reverse shock front is delayed in arriving at the Si/Ni
layer and there is little outward mixing of radioactive $^{56}$Ni.
These simulations include a 5\% perturbation in the velocity field.
Late spectral observations of the Type IIP SN 1999em show a very centrally
peaked H$\alpha$ line, although the peak is redshifted by $400\kms$
compared to the systemic velocity (Elmhamdi et al. 2003).
There is again strong observational evidence for H mixing to close
to the center.

The best studied case of a SN IIL/b is the Type IIb SN 1993J.
Because of the low H envelope mass, there is little deceleration
of the core material and numerical simulations show
relatively little mixing of the
H into the core region (Iwamoto et al. 1997).
The observed line profiles support this expectation; Houck \& Fransson
(1996) find that most of the H lies in the velocity range
$8500-10,000\kms$.
The O extends over the velocity $1000-4000\kms$, while the Fe extends
out to $3000\kms$ (Houck \& Fransson 1996), showing that mixing between
the inner core and the O region has occurred.
SNe Ib/c  have similarities to the SNe IIL/b in that
there is little or no H envelope so that H from the envelope is not
mixed to the center.
There is likely to be mixing of the heavy element core material.  From 
the [O I] line profile in the Type Ib SN 1985F, Fransson \& Chevalier
(1989) found that the distribution of O is  broader than would be expected
in a 1-dimensional model.
However, there may be lack of O inside of $1000 \kms$.
It should be kept in mind that the supernova types are not distinct, but there
is a continuous transition between the types.

Young PWNe have outer velocities $\sim 2000\kms$.
At this level, the expected composition for SN IIP and SN 1987A-like
events is a mixture of H envelope and core material.
In the other supernova types, the H envelope material is absent and
a mixture of core material is present.
Lower velocity material ($\la 500\kms$), relevant to the surroundings
of central neutron stars, depends on the mixing back of outer material.
The amount of mixing to these low velocities is uncertain in that the
reverse shock wave may lose power because of outer shock breakout
before the center is reached.
In SN 1987A and the Type IIP SN 1999em, mixing of H very close to the
center appears to have occurred.

\subsection{Fallback}

After an explosion is launched, some material can fall back to the central
neutron star (Colgate 1971; Chevalier 1989).
The nature of the fallback can depend on the supernova type.

Two types of fallback have been discussed.
In the first, an outward explosion is generated in a star but the inner
part of the flow falls back to the central object.
Numerical simulations of this process show that it is sensitive to the
energy and mass of the explosion (e.g., Herant \& Woosley 1994;
MacFadyen, Woosley, \& Heger 2001).
For a given energy, below a critical stellar mass,
there is little effect, but above that
mass, the effect becomes large (Herant \& Woosley 1994).

A sudden change in the possible fallback is indicated by the blast wave
self-similar solution
for a $\gamma=4/3$ flow in a medium with density profile $\rho=Dr^{-2}$
and a central mass $M_c$ (Chevalier 1989).
In this solution, there is a dimensionless constant, $\alpha=GM_cD/E$,
where $E$ is the energy.
There is a critical value of $\alpha_c=0.017$ below which the gas density
goes to zero at the center and above which the density $\rightarrow \infty$.
Allowing for a flow toward the neutron star because of neutrino losses,
it is clear that the solutions with a high central density will have
a much higher rate of fallback than the solutions with a low
central density.
Chevalier (1989) found that a $\rho=Dr^{-2}$ profile could approximate
the central density in SN 1987A and that the corresponding value of
$\alpha$ was less than, but close to, $\alpha_c$.
The sharp change is behavior is consistent with the numerical results.
If fallback occurs in this early phase, the material falling back is expected
to be that just outside the compact core.
If the object is in the strong fallback regime, the fallback is likely
to cause collapse to a black hole.
In the weak fallback regime, the amount of fallback is difficult to determine,
as it depends on the explosion mechanism.
Its composition is expected to be $^{56}$Ni, along with $^{4}$He.

The other mode of fallback involves matter that is brought back to
the center by the reverse shock wave (see \S~2.2).
Based on a  model for the central density in SN 1987A,
Chevalier (1989) estimated a fallback mass in the reverse shock phase
of $\sim 0.1\Msun$.
However, the recent simulations by Kifonidis et al. (2003) show a considerably
lower  density right at the center;
the simulations allow for accretion to a central compact object
and the low accretion rates imply an accreted mass $\la 0.001\Msun$
during the reverse shock phase.
These results show there is a large uncertainty in the fallback mass.
Any kick velocity of the neutron star could also affect its surroundings.

Chevalier (1989) suggested that the accreted mass from a normal
SN IIP would be lower than in the case of SN 1987A because the
more extended envelope would lead to accretion at a later time
when the central density is lower.
However, an examination of the central conditions at the time of
the interaction at the O/He interface (e.g., Shigeyama \& Nomoto 1990)
indicates that the accreted mass at that time can be comparable to
that occurring at the time of the He/H interaction.
In this case, there may not be significant differences in fallback
mass between 
the various core collapse supernovae except for the Type Ic supernovae
that have lost much of their He region and have low mass accretion.

The angular momentum of the accreted material is important for
the possible formation of a disk around the central object and
for spinning up the neutron star.
If the fallback material is from immediately around the neutron
star, the angular momentum per unit mass is likely to be similar to
that of the material going into the neutron star.
However, if material is mixed down to the central region in
connection with the reverse shock and Rayleigh-Taylor instabilities,
the outer material is likely to have a higher angular momentum per
unit mass, which can lead to disk formation.
Even if a disk forms, there may be substantial pressure support in
a disk that does not radiate efficiently, so we consider the accreted
material to have a rotational velocity of $\beta v_K$, where $v_K$ is
the Keplerian velocity.
If the neutron star rotates rigidly and is significantly
spun up, the rotation period after
the fallback accretion is
\begin{equation}
P_f\approx 4.6\left(\Delta M\over 0.1\Msun\right)^{-1}
\left(\beta\over 0.5\right)^{-1}
I_{45}  {\rm~ms}
\end{equation}
for a neutron star with $M=1.4\Msun$ and $R=10$ km,
where $I_{45}$ is the neutron star moment of inertia in units of $10^{45}$ g cm$^2$.
At the highest fallback mass estimates, the rotation of the
neutron star can be substantial.

The fallback of material is also responsible for the composition
on the surface of the remnant neutron star.
In the cases of SN IIL/b and SN Ib/c, H from the envelope is
not expected to be present, although material mixed down from
the He zone might be present.
In the case of SN IIP and SN 1987A-like, H from the envelope
might be accreted, depending on the efficiency of mixing downward
by the Rayleigh-Taylor instability and the velocity of the neutron star.

\subsection{Circumstellar Medium}
\label{seccsm}

Observations of supernovae at radio and X-ray wavelengths have given
us a fairly complete picture of the immediate surroundings of core
collapse supernovae (Chevalier 2003).
This information can be useful for the identification of supernova
types for young remnants because the initial interaction
is with this material.

SN 1987A has a complex and well-studied circumstellar medium resulting
from the late evolution to a RSG (red supergiant) and the subsequent evolution
back to a blue supergiant (McCray 2003).
The current interaction with the dense ring should continue for decades,
after which the supernova shock wave will be propagating in the slow, dense wind
from the red phase of evolution.
The outer radius of the RSG wind is probably limited by the
pressure, $p$, of the surrounding medium, and can be expressed as
(Chevalier \& Emmering 1989)
\begin{equation}
r_{RSG}=5.0\left(\dot M\over 5\times 10^{-5}\ml\right)^{1/2}
\left(v_w\over 15\kms\right)^{1/2}
\left(p/k\over 10^4~{\rm cm^{-3}~K}\right)^{-1/2} {\rm~pc},
\label{csrsg}
\end{equation}
where $k$ is Boltzmann's constant.
This outer boundary may have been observed as a shell
around SN 1987A (Crotts et al. 2001).

SNe IIP  end their lives as red supergiants with relatively low mass loss
rates.
Pooley et al. (2002) deduced a mass loss rate
$\sim (1-2)\times 10^{-6}\ml$ for $v_w=10\kms$ for the Type IIP SN 1999em
from X-ray and radio observations.
This rate of mass loss is consistent with expectations for single stars
of mass $\sim 10-15\Msun$ in the final phases of evolution (e.g.,
Schaller et al. 1992).
Because of the low mass loss rate, the RSG wind extends  to a relatively
small distance from the progenitor, $\la 1 $ pc (eq. [\ref{csrsg}]).
Outside of the RSG wind is   a low density wind bubble
created during the main sequence phase.

SNe IIL/b supernovae also
end their lives as red supergiants, but with higher mass loss
rates ($\ga 3\times 10^{-5}\ml$ for $v_w=15\kms$) than the SNe IIP.
The result can be a more extended dense circumstellar region,
extending to 5 pc or more from the star.
An interesting aspect of these supernovae is that the circumstellar
medium frequently shows evidence for CNO processing, with an enhanced
abundance of nitrogen.
Fransson et al. (2004) summarize the evidence for SN 1979C (IIL),
SN 1987A, SN 1993J (IIb), SN 1995N (IIn), and SN 1998S (IIL/n).
The mass loss has been sufficiently strong to reveal layers where
CNO processing has occurred.

SNe Ib/c are expected to have Wolf-Rayet star progenitors, which
have typical values of $\dot M \sim  10^{-5}\ml$ and $v_w\sim 10^3\kms$
in our Galaxy.
If the star had an earlier phase of RSG evolution, the dense wind
from the RSG phase is expected to be swept up by the fast wind.
For typical parameters, the mass loss rates for the the RSG and Wolf-Rayet
phases are comparable, but the Wolf-Rayet wind is 100 times faster
than the RSG wind.
The velocity of the swept up shell of RSG wind is expected to
be $10-20$ times the RSG wind velocity (Chevalier \& Imamura 1983),
or $100-200 \kms$.
The duration of the Wolf-Rayet phase spans a range, but a duration
$\sim 2\times 10^5$ yr is typical.
The expectation is that the RSG wind is completely swept up 
by the Wolf-Rayet wind; after the interaction shock breaks out into the
low density surrounding bubble, the swept up shell is subject to
Rayleigh-Taylor instabilities, as found in numerical simulations
(Garcia-Segura, Langer, \& MacLow 1996).
At the time of the supernova, the RSG wind material is in clumps
at a radial distance $\ga 10$ pc.
An overabundance of N is a possible signature of the circumstellar
origin of this material.

\subsection{Ionization at shock breakout}

Another aspect of the explosion that can depend on the supernova
category is the amount of ionizing radiation that is emitted at
the time of shock breakout.
The emitting surface area is the most important parameter for the
amount of radiated energy, so the radiative
effects of shock breakout are largest
for supernovae with red supergiant progenitors
(Klein \& Chevalier 1978).
For standard values of the opacity and a density parameter, the
amount of radiated energy at breakout for a red supergiant
explosion is (Matzner \& McKee 1999)
\begin{equation}
E_{rad}=1.7\times 10^{48}E_{51}^{0.56}
\left(M_{ej}\over 10\Msun\right)^{-0.44}
\left(R_*\over 3.5\times 10^{13}{\rm~cm}\right)^{1.74}
{\rm~ergs},
\end{equation}
where $R_*$ is the radius of the progenitor.
The strong dependence on $R_*$ is clear.
This expression should be applicable to both SNe IIP and
SNe IIL/b.
Because of mass loss, SNe IIL/b may typically have lower
values of $M_{ej}$ than the SNe IIP, which tends to give them
larger values of $E_{rad}$ because of the higher velocity of
the breakout shock wave.

If the initial radiation is degraded to ionizing photons with
an energy of $13.6\alpha$ eV (with $\alpha>1$), 
the number of ionizing photons is
$7.8\times 10^{58}/\alpha$ (for the reference values), which are
capable of ionizing a hydrogen mass of $65\Msun/\alpha$.
This shows that the ionizing radiation at the time of breakout
can plausibly ionize the mass loss from the progenitor star
in the red supergiant phase.
This is not the case when the mass loss is so dense that the
radiation dominated shock wave can be maintained in the
circumstellar medium.
Then the breakout occurs at such an extended radius that the
emission is at optical wavelengths and is non-ionizing
(e.g., Blinnikov et al. 2003).
However, this presumably applies to only a small fraction of
the red supergiant progenitors.

For a SN 1987A-like event, the reduced radius yields a number
of ionizing photons $(2-3)\times 10^{57}/\alpha$ (Lundqvist \& Fransson
1996; Matzner \& McKee 1999), which have the capability of
ionizing $\sim 2/\alpha \Msun$ of H.
In this case, the radiation may not be able to  ionize all of the
dense circumstellar mass.
The case of SN Ib/c progenitors is even more extreme because the
progenitor radius is $\sim 1\Rsun$ and the energy in ionizing
radiation is $\sim 10^{44}-10^{45}$ ergs.
In this case, only a small fraction of a $\msun$ can be ionized
and this mass is likely to be overrun early in the evolution of
the supernova remnant.

\section{PULSAR WIND NEBULA INTERACTION}

\label{pwn}

A first approximation to the PWN in hydrodynamic studies is that
it can be treated as a constant
pressure volume of $\gamma=4/3$ fluid (Ostriker \& Gunn 1971;
Reynolds \& Chevalier 1984; Chevalier \& Fransson 1992;
van der Swaluw et al. 2001).
In a toroidal
magnetic field model (Kennel \& Coroniti 1984) there can be
pressure gradients in the outer parts because of magnetic tension effects.
However, the stability of such a configuration is doubtful, and
polarization studies of PWNe indicate that the magnetic field does
not have a purely toroidal configuration.
In any case, the pressure at the outer contact discontinuity of the
nebula does not depend on the detailed properties of the nebula
(e.g., Bucciantini et al. 2003).

The basic equations for the evolution of the shell radius $R$,
velocity $V$, mass $M$, and interior pressure $p_i$ can be written
\begin{equation}
{dR\over dt}=V, \quad {dM\over dt}=4\pi R^2 \rho_{sn}\left(V-{R\over t}
\right)
\label{evol1}
\end{equation}
\begin{equation}
M{dV\over dt}=4\pi R^2\left[p_i- \rho_{sn}\left(V-{R\over t}
\right)^2\right]
\label{sw}
\end{equation}
\begin{equation}
{d(4\pi R^3p_i)\over dt}=L-p_i4\pi R^2{dR\over dt},
\label{evol3}
\end{equation}
where $\rho$ is the density in the freely expanding supernova ejecta
and $L$ is the power input from the central pulsar.
For the evolution of $L$, we make the standard assumption of evolution
with constant braking index $n$:
\begin{equation}
\dot E=\dot E_0 \left(1+{t\over \tau}\right)^{-(n+1)/(n-1)}.
\end{equation}
The vacuum dipole value of $n$ is 3, but observed values for pulsars
are found to be smaller: $2.51\pm 0.01$ for the Crab pulsar (Lyne et al. 1988);
$2.837\pm 0.001$ for PSR 1509--58 in MSH 15-52 (Kaspi et al. 1994);
$1.81\pm 0.07$ for PSR 0540--69 (Zhang et al. 2001);
$2.91\pm 0.05$ for PSR J1119--6127 in G292.2--0.5 (Camilo et al. 2000);
and $1.4\pm 0.2$ for the Vela pulsar (Lyne et al. 1996).
The fact that there is a range of values of $n$ indicates that the
assumption of evolution with constant $n$ is probably incorrect.
Any conclusions that depend on this assumption should be viewed
with caution.
We make the assumption here in order to investigate the effects of
pulsar power evolution and allow for various values of $n$.

As discussed above, the inner supernova density profile can reasonably
be described by a power law $\rho_{sn}=At^{-3}(r/t)^{-m}$.
Provided the pulsar is not a rapid rotator, the evolution of the nebula
takes place in this part of the supernova.
In this case, the evolution of the pulsar bubble is described by
the dimensional parameters $A$, $\dot E_0$, and $\tau$, and the
dimensionless parameters $n$ and $m$.
Characteristic quantities for the radius, velocity, shell mass, and
pressure can be found:
\begin{equation}
R_2=\left(\dot E_0\tau^{6-m}\over A\right)^{1/(5-m)}, \quad
V_2={R_0\over \tau}=\left(\dot E_0\tau\over A\right)^{1/(5-m)},
\end{equation}
\begin{equation}
M_2=(A^2 \dot E_0^{3-m}\tau^{3-m})^{1/(5-m)}, \quad
P_2=\left(A^3 \dot E_0^{2-m}\over \tau^{13-2m}\right)^{1/(5-m)}.
\end{equation}
These quantities can be used to nondimensionalize the basic variables
\begin{equation}
x={t\over\tau},\quad y={R\over R_2}, \quad w={V\over V_2}, \quad
z={M\over M_2}, \quad u={p_i\over P_2}.
\end{equation}
Substitution into eqs. (\ref{evol1}-\ref{evol3}) yields
\begin{equation}
{dy\over dx}=w, \quad {dz\over dx}=4\pi y^{2-m}x^{m-3}\left(w-{y\over x}\right)
\label{de1}
\end{equation}
\begin{equation}
z{dw\over dx}=4\pi y^2\left[u-y^{-m}x^{m-3}\left(w-{y\over x}\right)^2\right],
\label{de2}
\end{equation}
\begin{equation}
y^3{du\over dx}={1\over 4\pi(1+x)^{(n+1)/(n-1)}}-4y^2uw.
\label{de3}
\end{equation}

In the limit $x\ll 1$, the power input is constant and the evolution
can be solved analytically
\begin{equation}
y=B^{1/(5-m)}x^{(6-m)/(5-m)}, \quad w=\left(6-m\over 5-m\right)
(Bx)^{1/(5-m)},
\label{ans1}
\end{equation}
\begin{equation}
u={1\over 4\pi}\left(5-m\over 11-2m\right)B^{-3/(5-m)}x^{-(13-2m)/(5-m)}, \quad
z={4\pi\over 3-m} (Bx)^{(3-m)/(5-m)},
\label{ans2}
\end{equation}
where
\begin{equation}
B={(5-m)^3(3-m)\over 4\pi (11-2m)(9-2m)}.
\end{equation}
With this approximation,
the radius of the PWN  is
(Chevalier \& Fransson 1992; eq. [2.6])
\begin{equation}
R_p=\left[{(5-m)^3(3-m)\over (11-2m)(9-2m)}
{\dot E\over 4\pi A}\right]^{1/(5-m)}t^{(6-m)/(5-m)}.
\label{rad}
\end{equation}
For the particular case given in eq. (\ref{cden}) with $m=1.06$, we have
\begin{equation}
R_p=0.59 \dot E_{38}^{0.254}E_{51}^{0.246}
\left(M_{ej}\over 10\Msun\right)^{-0.50}
t_3^{1.254} {\rm~pc},
\label{rneb}
\end{equation}
where $t_3=t/(1000{\rm~yr})$.
A shock wave is driven into the freely expanding ejecta with
a velocity
\begin{equation}
v_{sh}={1\over (5-m)}{R_p\over t}.
\label{vsh}
\end{equation}
Another quantity of interest is the mass swept up by the PWN,
which can be written
\begin{equation}
M_{sw}={(5-m)^3\over (11-2m)(9-2m)}\dot E R_p^{-2} t^3.
\label{msw}
\end{equation}
The coefficient is only weakly dependent on $m$, and for $m=1.06$
is 1.0.
The internal energy in the bubble is
\begin{equation}
{E_{int}\over \dot Et} = {5-m \over 11-2m},
\end{equation}
which is 0.44 for $m=1.06$.

During the $x < 1$ phase, the swept-up shell is accelerated and is
subject to Rayleigh-Taylor instabilities (Chevalier 1977; Jun 1998),
which implies that after being shocked, the coupling between the
PWN and the ejecta is reduced.
In the extreme limit that there is no further acceleration of the
ejecta after it is shocked, the left-hand side of eq. (\ref{sw}) drops out; there
is direct pressure balance between the interior pressure and the
ram pressure of the shock front.
The solution for $y$, $w$, and $u$ is similar to that given in eqs.
(\ref{ans1}) and (\ref{ans2}),
except that now
\begin{equation}
B={(5-m)^3\over 4\pi (11-2m)}.
\end{equation}
Compared to the case where the ejecta are completely swept up,
$R$ is increased by a factor of 1.25 for $m=0$ and 1.37 for $m=1$.
This is for the extreme limit, and the actual factor by which the
radius is increased in likely to be smaller.
The value of $M_{sw}$ (eq. [\ref{msw}]) is increased by
the same factor as $B$, or 3 for $m=0$ and 3.5 for $m=1$.

Once $x>1$ ($t>\tau$), the power input from the pulsar drops
and the swept up material tends toward free expansion.
For $x\gg 1$, we have $V$ and $M_{sw}\rightarrow$  a constant value.
Consideration of eq. (\ref{de3}) shows that for $f\equiv (n+1)/(n-1) \ge 2$,
$p_i\propto t^{-4}$ and for $f < 2$, $p_i\propto t^{-(2+f)}$.
Results of integrating eqs. (\ref{de1}-\ref{de3}) are shown in
Fig. 1  for the case of a swept up shell.
The kinetic, $E_{kin}$, and internal, $E_{int}$,
energies are normalized to the initial rotation
energy of the pulsar, $E_{rot}$.
The final energy is larger than $E_{rot}$ because of the addition
of kinetic energy of the supernova gas.
Also shown is the instantaneous value of $\dot E t$, which may be
observed for some cases.
As expected, $\dot E t$ initially dominates, but becomes small for
$t>\tau$, especially for low values of $n$.
Measurements of $\dot E t$ and $E_{int}$ in a young PWN
can give an indication of the evolutionary phase of the pulsar.
We find that $E_{int}/\dot E t$ becomes large for $t/\tau >1$
and $(n+1)/(n-1) \ge 2$, and goes to a constant $>1$ for
$(n+1)/(n-1) < 2$.

The kinetic energy is less useful because the swept up mass is generally
difficult to observe and may not be swept into an outer shell
because of instabilities.
The evolution of $E_{int}$ can be found for the case of an unstable shell.
Eqs. (\ref{ans1}) and (\ref{ans2}) show that during the early phase,
$E_{int}\propto
R^3 p_i\propto y^3 u$
has no dependence on $B$, so that $E_{int}$ does not depend on whether
the stable or unstable case applies.
Integration of the differential equations to solve for the case $t>\tau$
shows that $E_{int}$ remains very close to the stable case, so that
the results given in Fig. 1 still apply.

These calculations can be used as follows when a PWN is still in
the inner part of the supernova density profile.
A lower limit on $E_{int}$ can be estimated from the synchrotron emission;
this limit occurs when there
is approximate equipartition between the particles
and magnetic field. 
With an estimate for the age $t$, the value of $\dot E t$ can be compared to
$E_{int}$.
A value of $E_{int}/\dot E t \ga 1$ implies that the pulsar has
undergone substantial spindown.
The value of $t/\tau$ can be estimated for a given value of $E_{int}/\dot E t$
for models with specified values of $m$ and $n$.
The results are not sensitive to the value of $m$ for a plausible range,
but they are quite sensitive to $n$; values of $t/\tau$ are smaller for
smaller $n$ over a plausible range (1.5--2.8).
Once $t/\tau$ is specified, the age and $\tau$ in the model are
\begin{equation}
t={2\over (n-1)}\left(t/\tau\over  1+t/\tau\right) t_{ch},  \qquad
\tau={2\over (n-1)}{t_{ch}\over  (1+t/\tau)},
\end{equation}
where $t_{ch}=P/(2\dot P)$ is the observed characteristic spindown age
of the pulsar.
The value of $t$ obtained this way must be checked for consistency with
the value assumed for $\dot E t$.
Once $t/\tau$ is determined, the initial values of $\dot E$ and $P$
are
\begin{equation}
\dot E_0=\dot E\left(1+{t\over \tau}\right)^{(n+1)/(n-1)}, \qquad
P_0=P\left(1+{t\over \tau}\right)^{-1/(n-1)}.
\end{equation}

For many observed PWNe, the only other piece of observational
data may be the radius of the nebula, $R$.
Using the value of $y$ corresponding to the value of $x$, we have
\begin{equation}
A={\dot E_0\tau^{6-m} y^{5-m}\over R^{5-m}}.
\end{equation}
This value of the central density parameter can be compared to the
value expected in a supernova.
For the case described by eq. (1), with $m=1.06$, we have
$A=1.3\times 10^{17}(M_{ej}/10 \msun)^{1.97}E_{51}^{-0.96}$, where the
numerical factor is in cgs units.
Comparison with the value derived from the nebular model provides
a consistency check on the model.

These considerations only apply while the PWN is in the inner, flat
part of the supernova density profile.
The time to reach the transition velocity, assuming constant
power input, can be found by combining
eqs. (\ref{vtr}) and (\ref{rad}) and setting $v_{tr}=R_p/t$:
\begin{equation}
t_{tr}={2(b-5)(11-2m)(9-2m)\over (3-m)(5-m)^2(b-m)}{E\over \dot E_0}.
\end{equation}
For $b=9$ and $m=1$, we have $t_{tr}=1.97E/\dot E_0$.
In the case where a shell is not swept up because of
instabilities, the bubble expands more
rapidly and $t_{tr}=0.56E/\dot E_0$ for $b=9$ and $m=1$.
In either case, the condition is approximately $\dot E_0t_{tr}\approx E$,
or the energy injected by the pulsar must approximately be that
of the expanding supernova gas; the energy required to displace the
supernova gas is approximately the kinetic energy in the ejecta.
The condition that the bubble reach the bend in velocity is thus
that the initial pulsar rotational energy be greater than the
kinetic energy in the supernova, or $I\Omega_0^2/2 > E$,
where $I$ is the moment of inertia of the pulsar and $\Omega_0$ is
its initial spin rate.
This condition can be written as
\begin{equation}
P_0 < 9 I_{45}^{1/2}E_{51}^{-1/2}\quad {\rm ms},
\end{equation}
where $I_{45}$ is the moment of inertia in units of $10^{45}$ g cm$^2$.

If a PWN were able to expand past the inflection point in the
supernova density profile where the profile becomes steeper
than $\rho\propto r^{-5}$, there is a substantial difference in
the evolution depending on whether the swept-up shell breaks
up by Rayleigh-Taylor instabilities.
If the shell remains intact, the evolution is determined by
the acceleration of a shell of fixed mass, with the
result $R\propto t^{1.5}$ (Ostriker \& Gunn 1971).
However, if the shell breaks up, the expansion is determined
by pressure equilibrium, with $R\propto t^{(6-m)/(5-m)}$.
It can be seen that as $m\rightarrow 5$ or higher, the bubble
expands rapidly into the low density medium and is expected
to move out to the place where the outer parts of the supernova
are interacting with the surrounding medium.
In view of the expected Rayleigh-Taylor instability of the
accelerated shell, the blow out scenario appears to be the
most plausible (see also Bandiera, Pacini, \& Salvati 1983).

In this section, we have neglected possible sources of energy
loss from the pulsar and PWN, such as gravitational radiation and
radiative losses.
For the initial periods found in \S~5, gravitational radiation
is not important.
Radiative cooling may be important for the PWN at early times,
but it is uncertain because it depends on the energy distribution
of injected particles and the evolution of the distribution.
The losses do not directly affect the magnetic field, which could
drive the bubble expansion;  however, a highly magnetized bubble
may be subject to instabilities and magnetic reconnection.
There may be an observational test of strong radiative losses.
For a radiative phase lasting $\la 10^3$ yr, the radiative luminosity would
be $\ga 6\times 10^{39}\ergs$ in order for the initial period
to be $<10$ ms.
This luminosity should be accessible in supernova observations, but
steady luminosities of this order have not been reported, so that
rapid initial rotation cannot be accomodated in this way.

There may be other modes of neutron star spindown at early times;
however, if the injected power does work on the inner edge of the supernova
ejecta, as assumed here, the process will affect the appearance
of the PWN.

\section{CIRCUMSTELLAR INTERACTION FOR SN IIL/b}

As discussed in \S~\ref{seccsm}, the circumstellar
environment of massive stars is
shaped by their mass loss and can be complex.
In general, specific hydrodynamic models are needed to follow the mass
loss interactions in the presupernova stage and the subsequent
supernova (e.g., Dwarkadas 2001).
Detailed models for individual supernova remnants are problematical because
of the many parameters.
One case that is more amenable to analysis is the case of a SN IIL/b
interacting with the extended dense wind from the progenitor star.
For the density profile of the exploded star, $\rho_{sn}$,
we take the radiative star
explosion model of Matzner \& McKee (1999), with the profile in the
harmonic mean approximation; eq. (1) represents the inner part of this density
profile.
The external medium is taken to be a freely expanding wind, with
$\rho_{cs}=\dot M/4\pi r^2 v_w\equiv Dr^{-2}$.
We define $D_*=D/1.0\times 10^{14}{\rm~g~cm^{-1}}$, where the
reference value corresponds to 
$\dot M= 3\times 10^{-5}\ml$ and $v_w=15\kms$.
The circumstellar mass swept up to $R$ is
$M_{sw}=9.8 D_* (R/5{\rm~pc})\Msun$.

In the thin shell approximation, the equations for the evolution
of the shell radius $R$,
velocity $V$, and mass $M_s$ are (Chevalier 1982)
\begin{equation}
{dR\over dt}=V, \quad {dM\over dt}=4\pi R^2 \left[\rho_{sn}\left({R\over t}-V
\right)+\rho_{cs}V\right]
\label{evol5}
\end{equation}
\begin{equation}
M{dV\over dt}=4\pi R^2\left[\rho_{sn}\left({R\over t}-V
\right)^2 -\rho_{cs}V^2\right].
\label{evol6}
\end{equation}
The parameters of the problem are $E$, $M$, and $D$, so dimensionless variables
can be introduced:
\begin{equation}
x={t\over t_1},\quad y={R\over R_1}, \quad w={V\over V_1}, \quad
z={M\over M_1},
\end{equation}
where
\begin{equation}
R_1={M_{ej}\over D},\quad V_1=\left(E\over M_{ej}\right)^{1/2},
\quad M_1=M_{ej}, \quad t_1={R_1\over V_1}={M_{ej}^{3/2}\over D E^{1/2}}.
\end{equation}
Substitution into eqs. (\ref{evol5}-\ref{evol6}) yields
\begin{equation}
{dy\over dx}=w, \quad {dz\over dx}=4\pi
\left[{\rho_{sn}\over DR_1^{-2}}\left({y\over x}-w\right)y^{2}+w\right],
\label{de4}
\end{equation}
\begin{equation}
z{dw\over dx}=4\pi
\left[{\rho_{sn}\over DR_1^{-2}}\left({y\over x}-w\right)^2y^{2}-w^2\right],
\label{de5}
\end{equation}
where
\begin{equation}
{\rho_{sn}\over DR_1^{-2}}=0.0432 x^{-3}\left[\left(w\over 2.30\right)^{0.236}
+\left(w\over 2.30\right)^{2.261}\right]^{-4.5}.
\end{equation}
The initial part of the evolution is dominated by the outer steep power
law part of the supernova density profile.
This part of the evolution is self-similar, with
$y_{ss}=R_{ss}/R_1=1.314 (t/t_1)^{0.878}$.
The self-similar solution provides the initial conditions for
the integration of eqs. (\ref{de4}-\ref{de5}).

The results of integrating eqs. (\ref{de4}-\ref{de5}) are shown in Fig. 2, where
the deceleration parameter for the shell, $s$, is defined by
$s=Vt/R$, $M_{sej}$ is the ejecta mass swept into the shell,
and $M_{scs}$ is the circumstellar mass swept into the shell.
It can be seen that the solution gradually evolves away from the early
self-similar solution to a case with $s=0.5$.
This value of $s$ is expected for the thin shell approximation, which
is equivalent to the assumption of radiative shocks and momentum
conservation at late times.

The thin shell approximation breaks down in the energy conserving
case as the shocked region becomes broader; this transition
is expected once the reverse shock front propagates back into
the flat part of the supernova density profile.
For an energy-conserving blast wave in a wind, the forward shock
wave expands 
\begin{equation}
R=(3E/2\pi D)^{1/3}t^{2/3}=5.4E_{51}^{1/3}
D_*^{-1/3}t_3^{2/3}{\rm~ pc},
\end{equation}
where $t_3=t/({\rm 1000~yr})$.
In terms of the variables used in this section, we have $y=0.7816 x^{2/3}$.

Typical values for the parameters are $E_{51}=1$, $D_*=1$,
and $M_{ej}=5\Msun$, leading to $R_1=32.1$ pc, $V_1=3160\kms$,
and $t_1=9930$ yr.
With these parameters, Fig. 3 shows results for the thin shell model
(solid line) and the blast wave model (dashed line) over a radius-time
range of interest.
The blast wave case does not explicitly depend on $M_{ej}$ but requires that
the remnant be in an evolved state because of a low value of $M_{ej}$
compared to the mass lost in the wind.

\section{COMPARISON WITH OBSERVED PULSAR WIND NEBULAE}

A list of probable young PWNe and supernova remnants
in which central pulsars have
been identified is given in Table 1; the list includes all of the
objects, estimated to have an apparent age $<5$ kyr, 
in Table 2 of Camilo et al. (2002a), except for N157B.
One addition is the recently discovered pulsar in G54.1+0.3
(Camilo et al. 2002b).
These objects are plausibly
interacting with ejecta; N157B has an asymmetric morphology that
suggests the PWN has interacted with the reverse shock front and
is not considered here.
References to the Galactic pulsar and remnant properties can
be found in Green (2004).
A recent reference to 0540--69 in the Large Magellanic Cloud is
Hwang et al. (2001).
The fourth column gives the current spindown power of the pulsar,
$\dot E=4\pi^2 IP^{-3}\dot P$, where $I$, the neutron star
moment of inertia, is given in terms of $I_{45}=I/10^{45}{\rm~g~cm^2}$.
The fifth column gives the observed pulsar period, $P$, and the
sixth column gives the characteristic pulsar age, $t_{ch}=P/2\dot P$.
If the pulsar is born rotating much more rapidly than the
current rate and the braking index is $n=3$, then $t_{ch}$
is the actual age.
If the pulsar is born with a period close to its current
period, it can be younger than $t_{ch}$.
Alternatively, if the pulsar is born spinning rapidly and
has a braking index $n<3$, it can be older than $t_{ch}$.
The next two columns give the radii of the pulsar wind nebula and
of the surrounding supernova remnant, if present.
In some cases, the nebulae are asymmetric so that the quoted radius
is a mean value.

As discussed in \S~3, the internal energy in the PWN, $E_{int}$, can
provide information of the evolutionary state of the nebula.
The minimum internal energy, $E_{min}$, can be found from the synchrotron emission.
PWNe typically have a moderately flat radio spectrum, with electron energy
index $p_1<2$ and a steeper X-ray spectrum with energy index $p_2>2$.
At some intermediate energy $E_b$, there is a break in the spectrum, at which the
particles radiate at frequency $\nu_b$.
Most of the particle energy in the nebula is in particles with energies near $E_b$.
Following Pacholczyk (1970), the energy in particles with $E<E_b$ is
\begin{equation}
E_{p1}={2 c_1^{1/2}\over c_2(2-p_1)}\,\nu_b^{1/2}B_{\perp}^{-3/2}L_{\nu b},
\end{equation}
where $c_1=6.27\times 10^{18}$ and $c_2=2.37\times 10^{-3}$ (cgs units) are
constants used in Pacholczyk (1970), $B_{\perp}$ is the perpendicular magnetic
field, and $L_{\nu b}$ is the spectral luminosity at $\nu_b$.
For a spectrum that is continuous across the break and that extends to high
and low frequencies, we have $E_{p2}/E_{p1}=(2-p_1)/(p_2-1)$, where $E_{p2}$
is the energy in particles above the break, leading to the total particle energy,
$E_p=E_{p1}+E_{p2}$.
The minimum energy condition is $E_B=(3/4)E_p$, where $E_B$ is the magnetic
energy.
The minimum total energy ($E_p+E_B$) is then
\begin{equation}
E_{min}=1.0\times 10^7\left({1\over 2-p_1}+{1\over p_2-2}\right)^{4/7}
 \nu_b^{2/7}R^{9/7}L_{\nu b}^{4/7} {\rm~ergs},
\end{equation}
where cgs units are used.
Estimates of $E_{min}$ for the PWNe are given in Table 2, where the input
parameters are also given.
A related, but more complicated, expression for $E_{min}$ is given in
Blanton \& Helfand (1996); they found $E_{min}=9.9\times 10^{47}$ ergs for
Kes 75, which is in good agreement with the value deduced here.
The results are  sensitive to the distance, $d$, with
$E_{min}\propto d^{17/7}$.
Also, there may more than one break between radio and X-ray wavelengths,
or a more gradual turnover, which would reduce the value of $E_{min}$.
As more observations become available on the spectra of PWNe, the energy
estimates can be improved.

\subsection{Crab Nebula}

The abundances in the Crab Nebula point to an initial progenitor star in
the $8-10\Msun$ range because of the lack of an O-rich mantle
(Nomoto et al. 1982).
This is likely to be in the range of the Type IIP supernovae, as indicated
by recent observations of supernova progenitors (\S~1).
For this type of supernova, core material is decelerated during
the explosion and Rayleigh-Taylor instabilities mix H envelope gas
to low velocities, as is observed in the Crab Nebula.
The expansion of the pulsar nebula in a medium like
that described by eq. (1) gives results that are in reasonable
agreement with the observed properties of the Crab Nebula (Chevalier
1977; Chevalier
\& Fransson 1992).

As discussed in \S~2.4, SNe IIP may have relatively weak winds during
the progenitor red supergiant phase.
The current radius of the Crab Nebula, 2 pc, is larger that the expected
extent of the RSG wind and the forward shock should be at a considerably
larger radius.
Since the extent of the RSG mass loss is likely to be
$\la 1$ pc, the forward shock is currently in the
low density surrounding bubble.
Because the dominant source of X-ray emission is likely to be from
gas that passed through a reverse shock at an earlier time and is
now undergoing adiabatic expansion, detailed simulations are needed
to show whether the expected emission would fall below observational
limits.
The ultimate test of models for the Crab is the detection of
freely-expanding supernova
gas or an interaction region outside of the well-observed PWN.
H$\alpha$ emission is expected from the freely expanding ejecta
and current upper limits are getting into an interesting range
(Fesen et al. 1997).

\subsection{3C 58}

Of the historical supernovae prior to 1500 AD, SN 1054 and SN 1006
have the most secure identifications with supernova remnants;
in addition to positional agreement, the expansion of the remnants
is consistent with the implied ages (Stephenson \& Green 2002).
Stephenson \& Green (2002) also consider the identification of SN 1181
with 3C 58 to be secure, based on the length of time of Chinese
observations, the position of the supernova and
the lack of other viable candidates for the remnant.
However, in this case the radio (Bietenholz, Kassim, \& Weiler 2001)
and optical (Fesen, Kirshner, \& Becker 1988)   expansion suggest an
age greater than 820 years, although the smaller age is not ruled out
if rapid deceleration has occurred.

An alternative point of view is that the remnant should not
be identified with SN 1181 and is actually  older.
Chevalier (2004) gave arguments for this being the case, based
on the PWN.
These arguments were based on models with constant power pulsar period because
$t_{ch}=5390$ yr for 3C 58, which is much larger than the age of SN 1181.
One argument is that currently $\dot Et < E_{int}$, implying
that the pulsar has significantly spun down, which is inconsistent
with the small age.
Another is that the rapid expansion of the PWN requires  lower
density surrounding ejecta than can plausibly be expected
for a supernova.
Finally, the expected mass swept up by the PWN is smaller than
that inferred from X-ray observations.
With the small age, eq. (\ref{msw}) can be used, yielding
$M_{sw}=0.002\Msun$; the observations indicate
$M_{sw}\sim 0.1\Msun$ (Bocchino et al. 2001) or
$\ga 0.5\Msun$ (Slane et al. 2004).

Another constraint comes from the temperature of the X-ray emission,
which is found to be $kT=0.23$ keV (Bocchino et al. 2001; Slane et al. 2004).
The shock velocity required to produce this temperature is
$v_{sh}=343 \mu^{-1/2}\kms$, where $\mu$ is the mean particle weight per amu.
Assuming a composition that is half H and half He by mass, $\mu=1$.
If 3C 58 is identified with SN 1181 and $m=1.06$, eq. (\ref{vsh}) yields
$v_{sh}=1000\kms$, which would substantially overheat the shocked gas.

A first approximation for a model with a larger age is to assume steady power
injection and take an age $t=2400$ yr.
Then $E_{min}/\dot E t=0.49$ and $v_{sh}=341\kms$, in accord with expectations
provided there is no strong deviation from energy equipartition.
The swept up mass from eq. (\ref{msw}) is $0.056\Msun$.
This is lower than observed, but the situation is improved by
considering the case of no ejecta acceleration after the shock wave passes;
$M_{sw}$ is increased to $0.2\Msun$, in approximate accord with observations.
In this model, the shell radius is $R_p=1.74 E_{51}^{0.246}
(M_{ej}/10\Msun)^{-0.50}$ pc, which is somewhat lower than observed.
The assumption of steady power is inaccurate because $t$ is
not much smaller than $t_{ch}=5390$ yr.
Consideration of evolutionary models shows that smaller ages are
preferred by $v_{sh}$, but larger ages are preferred by $M_{sw}$ and $R_p$.
Overall, the indications are that $t=2400\pm 500$ yr and $P_0\approx 50$ ms.

The main problem with an age of $2400\pm 500$ yr for 3C 58 is where, then,
is the remnant of SN 1181; there are no plausible candidates
at the expected position.
However, some remnants with ages $\sim 10^3$ yr may be difficult
to detect.
The emission we observe from the Crab Nebula is essentially all due
to the power from the Crab pulsar.
If the central compact object were a low power pulsar or an
object like that in Cas A, the supernova remnant could
remain undetected.

With an age of several 1000 yr, the optical
filaments and knots can be ejecta that
have been overrun by the PWN.
The knots are then expected to overlap the PWN, as observed.
The presence of relatively slow, H-rich ejecta 
(Fesen, Kirshner, \& Becker 1988) then suggests
a SN IIP progenitor, as in the case of the Crab.
The lack of observed interaction around 3C 58 can also be attributed
to the low mass loss for this case.

\subsection{0540--69}

The PWN in this object has already been examined in the
context of interaction with freely expanding supernova ejecta
(Reynolds 1985; Chevalier \& Fransson 1992).
Although the free expansion age for the nebula around the pulsar
is $690\pm 45$ years for a distance of 50 kpc (Kirshner et al. 1989),
acceleration of the nebula by the pulsar bubble  yields a
somewhat larger age, $\sim 830$ years.
The age is considerably less than the characteristic age of the
pulsar, 1660 yr, so the nebula should be in the early stage of evolution.
This is borne out by the ratio $E_{min}/\dot E t=0.13 (t/800{\rm~yr})^{-1}$.
The ratio $E_{int}/\dot E t=0.44$ is attained if $E_{int}$ is several
times the minimum energy.
A realistic model must allow for some evolution; a model like that
described in \S~3 with $E_{int}/\dot E t=0.58$, $m=1.06$, and $n=2.0$ 
yields $t=790$ yr, $\tau=2530$ yr, $\dot E_0=3.4\times 10^{38}\ergs$,
$P_0=38$ ms, and $A=2.2\times 10^{16}$.

The remnant shows a larger shell in radio and X-ray emission,
with some optical emission.
The radio shell has a diameter of 17.5 pc (Manchester, Staveley-Smith, \&
Kesteven 1993),
leading to a mean velocity of $10,300\kms$ and implying that only
high velocity ejecta have been decelerated by the surrounding medium.
Despite the high mean velocity, the temperature of the X-ray
emission is $\sim 4$ keV and the abundances are normal for
LMC material (Hwang et al. 2001).
These properties are consistent with the shock wave moving
into  a dense circumstellar
shell.
The presence of optical emission is indicative of dense clumps within
the shell.
As noted in \S~2.4, a shell with these properties can be formed around
a Wolf-Rayet star, which ultimately explodes as a SN Ib/c.
If the shell is composed of material from a previous RSG phase,
an overabundance of N might be expected.
Mathewson et al. (1980) note that there is an arc of emission to the
SW of the pulsar with strong [NII] emission, although the relation
of the emission region to the supernova remnant is not completely
clear.

In a SN Ib/c scenario, slow moving ejecta around the PWN should be
free of H.
The presence of H$\alpha$ emission in the inner nebula has been
controversial (Dopita \& Tuohy 1984; Kirshner et al. 1989), but
the absence of H$\beta$ emission is consistent with the H-free
hypothesis.
The absence of H$\alpha$ emission in the inner nebula is a
prediction of the  scenario proposed here.

\subsection{Kes 75}

The pulsar in Kes 75 is notable for its short characteristic
age, $P/2\dot P=723$ years; however, the actual age is  uncertain
because of the uncertain amount of spindown and the uncertain
braking index (Helfand, Collins, \& Gotthelf 2003).
There are signs that the pulsar has undergone substantial spindown.
One is that currently $\dot Et=2.5\times 10^{47}t_3$ ergs, where
$t_3$ in the age in $10^3$ yr, whereas the minimum energy in fields
and particles in the synchrotron nebula is $1\times 10^{48}$ ergs
(Blanton \& Helfand 1996; Table 2).
We have $E_{int}/\dot Et\approx 4$ if the particles and fields
are close to equipartition and higher if not.
Figure 1 shows that this is an indication of an evolved pulsar.
Another indication is the size of the PWN.
The size of the synchrotron nebula in Kes 75 is comparable to that in
0540--69, although the current $\dot E$ is smaller by a factor of
18.

The model described in \S~3 has been used to
search for a consistent model for Kes 75, assuming the inner
supernova density profile given by eq. (1).
For $n$ in the range $2.0-2.8$, we found a set of models with
$E_{int}/\dot Et\approx 4$ that yield $A\approx 1\times 10^{17}$ (within a
factor of 2), which is consistent with expansion in a normal
supernova.
The value of $P_0$ in these models  is $\sim 30$ ms.
Other model results show more variation with
$t=1300, 940, 800$ yr, $\tau= 140, 27, 8.4$ yr,
and $\dot E_0=8.6\times 10^{39}, 3.3\times 10^{40}, 1.3\times 10^{41}
\ergs$, for $n=2.0, 2.5, 2.8$, respectively.

Kes 75 has an outer shell structure at X-ray and radio wavelengths
that is similar to that observed in 0540--69 (Helfand et al. 2003).
The shell radius is 9.7 pc, so that the mean expansion velocity
is $9,500 t_3^{-1} \kms$.
As in the case of 0540, the large mean velocity is suggestive
that the supernova has traversed a region of low density and is
now interacting with a higher density shell.
These properties are indicative of a SN Ib/c.

\subsection{MSH 15-52}

This remnant is notable for its large size relative to its age.
Radio emission to the SE of the pulsar is at a distance of
25 pc from the pulsar (for a distance of 5.2 kpc, Gaensler
et al. 1999).
For an age of 1700 yr, the characteristic spindown time of
the pulsar, the mean velocity to this region is
$14,000\kms$, suggesting that the supernova ejecta have traveled
relatively unimpeded to this region.
Gaensler et al. (1999) argue for a high $E/M$ ratio and a
SN Ib/c origin.
This supernova type is plausible and we  build on that scenario here.

The PWN itself is large and of low surface brightness in X-rays, and is not
clearly observed at radio wavelengths.
With the current $\dot E=1.8\times 10^{37} I_{45}\ergs$,
we have $\dot Et=9.7\times 10^{47} I_{45} (t/1700 {\rm ~yr})$ ergs,
which can be compared to the minimum energy estimate of $1.5\times 10^{48}$ ergs
(Table 2).
This estimate is quite uncertain because of the uncertain radio observations
and the possibility that the spectrum flattens just below X-ray
wavelengths (Gaensler et al. 2002).
The additional constraint of expansion into a plausible supernova
surrounding medium shows that significant evolution (i.e. spindown)
of the pulsar is needed.
The point is that considerable energy is needed for the PWN boundary to have
expanded to such a large size and fairly rapid initial rotation
of the pulsar is indicated, unless the surrounding supernova has
extremely low mass or high energy.
As an example, a model with $n=2.8$ and $E_{int}/\dot Et = 4$
yields $t=1870$ yr, $\tau=19$ yr, $E_0=3\times 10^{41} \ergs$,
$P_0=12$ ms, and $A=3\times 10^{16}$ in cgs units.
The internal energy is $2-3$ times the equipartition
value.
This model is comparable to that proposed in Chevalier \& Fransson (1992).

The optical knots that comprise RCW 89 (the optical counterpart
of MSH 15-52) are at a distance $\ga 10$ pc from the
pulsar and are presumably emitting by supernova interaction.
In the SN Ib/c explosion scenario, they are possibly material in
the RSG wind that was swept up during the Wolf-Rayet phase
(\S~4).
The strong [NII] lines from this gas (Seward et al. 1983) provide
support for this identification.

The model for the PWN gives an age that is consistent with the
remnant originating in SN 185.
In particular, the evidence for strong spindown and the braking
index near 3 imply an age within about 200 yr of $t_{ch}=1700$ yr.
The combination of large PWN and small age implies that the
ejecta should have been heated to X-ray emitting temperatures
by the expanding PWN.

\subsection{G292.0+1.8}

At a distance of 6 kpc (Gaensler \& Wallace
2003), the 8$^{\prime}$ diameter of the remnant corresponds to
a radius of 7 pc.
The presence of O-rich, H-poor filaments at a velocity of $\sim 2000\kms$
(Murdin \& Clark 1979)
immediately places the supernova in the IIL/b or Ib/c categories.
The interaction is presumably with the RSG wind of the progenitor
star, and the current radius indicates that the shock front is in
the outer parts of the wind, implying a Type IIL/b identification.
The structure that is observed in the outer X-ray emission, such as
the filament of emission across the E--W direction (Park et al. 2002) 
is probably structure in the RSG wind; 
this is consistent with the
normal abundances found in this region.

The displacement of the PWN from the center of the supernova remnant
can be interpreted as implying a kick velocity of the neutron star
of $\sim 480\kms$ for a distance of 6 kpc and an age of 3200 yr
(Hughes et al 2001).
Alternatively, the displacement may be the result of an asymmetric
interaction with the circumstellar medium, as appears to be the
case with Cas A (Reed et al. 1995).
In either case, the pulsar is surrounded by uniformly expanding supernova
ejecta, so that the models of \S~3 apply, provided that there is not
a strong radial density gradient in the ejecta.
Taking a distance of 6 kpc and $R=3.5$ pc for the PWN (Gaensler \&
Wallace 2003), we have calculated a number of models with $n=2.0,2.5,2.8$.
In order to obtain $A\approx 10^{17}$, the models have
ages of $2700-3700$ yr, $P_0=30-45$ ms, and $\dot E_0=(3-30)
\times 10^{38}\ergs$.

The internal energy in the models is $(1-2)\times 10^{48}$ ergs,
which can be compared to the minimum energy of 
$E_{min}=1\times 10^{48}$ ergs (Table 2).
Hughes et al. (2003) estimate an internal energy $\sim 4\times 10^{49}$ ergs,
based on the assumption that the similar extent of radio and X-ray
emission implies that the synchrotron lifetime of the particles is
comparable to the age.
The results of \S~3 show that the kinetic energy of the nebula would
have to be much larger than $E_{int}$ when $E_{int}/\dot Et>30$.
For $n=2$, the value of $E_{kin}$ would be $\sim 4\times 10^{51}$ ergs
and would be larger for larger values of $n$.
This energy is much larger than that inferred for any other
PWN, and the size of the
PWN would be larger than observed for any plausible age and supernova density.
These considerations suggest that the value of $E_{int}$ is closer
to the equipartition value and that the X-ray extent is not an
indicator of particle lifetimes.

The strong interaction that the remnant shows out to the outer
shock wave at $R=7$ pc can be interpreted in terms of interaction
with the dense wind expected to be present around a Type IIL/b
supernova.
The radius is close to the maximum extent that is expected for
such a wind (\S~2.4).
Ghavamian, Hughes, \& Williams (2004) have detected clumpy  gas that is
superposed on the remnant and is probably photoionized.
This material may  
have been photoionized by the radiation at shock breakout, although
ionization by radiation from the supernova remnant is also a possibility.
The X-ray luminosity suggests a swept up mass of $\sim 7-8\Msun$
(Gonzalez \& Safi-Harb 2003a), which is consistent with a RSG wind.
For the reference wind parameters used before ($\dot M=3\times 10^{-5}\ml$
and $v_w=15\kms$), the swept up mass is $13.7 (R/7{\rm~pc})\Msun$.
The wind interaction models of \S~4 can be used to test the
radius and age for consistency.
Fig. 3 shows that the remnant's properties are roughly consistent
with those expected in the interaction models.

The fact that heavy element rich knots with velocities $\ga 1000\kms$
have been observed (Murdin \& Clark 1979) supports the
Type IIL/b identification.
However, the age of $\sim 3200$ yr deduced in the
present model is greater than the
$\la 1600$ yr estimated by Murdin \& Clark (1979) from the knot velocities.
Murdin \& Clark obtained their limit by dividing the observed velocity
range ($2030\kms$) into the 3.0 pc diameter that they estimated 
for the remnant.
The emitting knots could actually be within the 14 pc diameter of the
outer shock front (Gaensler \& Wallace 2003), which gives an age
limit of $<6700$ yr.
This limit can be reduced by considering the asymmetry in the line profiles
and projection effects; a study of the fast optical knots has the
potential to yield a realistic age estimate for the remnant.
After this paper was submitted, the results of Ghavamian et al. (2004)
on the kinematic age became available.
They found an age of 3400 years, in good agreement with the age
found here by an independent method.

\subsection{G11.2--0.3}

This remnant is another excellent case of both a PWN and circumstellar
shell interaction.
At a distance of 5 kpc (Green et al. 1988), the radius of the remnant is 3.3 pc,
so the RSG wind is an immediate candidate for the interaction,
which places the remnant in the Type IIL/b category.
Green et al. (1988) find that the radio emission does not have a well-defined
outer edge, so the outer shock front could be at a larger radius.
The PWN is not symmetric, but its radius can be approximated as
0.9 pc (Tam, Roberts, \& Kaspi 2002).
Clark \& Stephenson (1977) suggested that G11.2--0.3 is the remnant
of SN 386.
Stephenson \& Green (2002) note that the information on this event
does not conclusively identify it as a supernova and that there
are other candidate supernova remnants; however, it appears to be
the remnant in the region with the smallest diameter and largest surface
brightness.
The identification with SN 386 and the remnant's small 
size imply an age much less
than the pulsar spindown age of 24,000 yr (Torii et al. 1999), 
which suggests that the
current pulsar period and power are close to their initial values.

Taking the age to be 1618 yr, we have $\dot E t=3.3\times 10^{47}I_{45}$ ergs,
to be compared to the minimum energy in particles and fields 
$E_{min}\approx 3\times 10^{46}$ ergs (Table 2).
With a steady input of energy, we expect $E_{int}/\dot E t=0.45$ and,
in this case, $E_{min}/\dot E t=0.1$, so there in no problem in the pulsar
producing the internal energy in the PWN.
Some deviation from equipartition is indicated; the alternative is
that the remnant is younger than SN 386.
Substituting the age and $\dot E$ into eq. (28) yields
$R_{PWN}=0.54E_{51}^{0.246}(M_{ej}/10\Msun)^{-0.5}$ pc.
Compared to the observed value of 0.9 pc, a small value of the ejected mass
is indicated, although uncertainties in the model (due to, e.g., the
asymmetry of the nebula) do not make it possible to conclude more than
that the observations are roughly consistent with model expectations.

Fig. 3 shows that the small size of G11.2--0.3 is consistent with
an origin in SN 386 if it is compared to the blast wave model.
In this case, the deceleration parameter is $m\approx 0.66$
at the forward shock front;
the reverse shock is expected to have
a lower value of $m$ than the forward shock.
From radio observations, Tam \& Roberts (2003) deduced
$m= 0.68\pm 0.14$ (20 cm observations) and
$m= 0.48\pm 0.16$ (6 cm observations).
The agreement with the model value is adequate.

\subsection{G54.1+0.3}

This object has the interesting feature that the pulsar has similar
properties ($P$ and $\dot P$) to the one in G292.0+1.8, but it
is lacking the surrounding circumstellar interaction.
There are also differences in the PWNe, with G54.1+0.3  being
considerably less luminous at radio wavelengths
and having a smaller size.
Considering that other PWNe appear to have $E_{int}$ close to $E_{min}$,
that $E_{min}=8\times 10^{46}d_5^{17/7}$ ergs (Table 2)
 where $d_5$ is the distance in units of 5 kpc,
and that $\dot Et=3.8\times 10^{47}I_{45}t_3$ ergs, the indications are that
G54.1+0.3 is in an early evolutionary phase, with an age $< \tau_{ch}=2900$ yr.
Substituting $\dot E_{38}=0.12$ and $M_{ej}=5\Msun$ in eq. (28) yields
$R=0.5t_3^{1.254}$ pc for the swept up shell case and $R=0.7t_3^{1.254}$ pc 
for the unstable case.
$\dot E_{38}$ should be somewhat larger than 0.12 because of evolution,
but this does not substantially change the results.
The indications are that $t\approx 1500$ yr and $P_0\approx 100$ ms,
using the relations from \S~3.
This examination of the PWN shows that although this pulsar and the
one in G292.0+1.8 have similar magnetic fields, the one in G54.1+0.3 was born
with a significantly larger period and is younger.
The differences between the PWNe cannot be attributed solely to different
supernova surroundings because the ratio $E_{int}/\dot Et$ depends only
on $t/\tau$ and $n$, and not on the supernova properties.

Given the age of G54.1+0.3 and the lack of strong circumstellar interaction,
a Type IIL/b supernova is ruled out.
Beyond that, there are at present no further clues on the supernova type.
The similarity to the Crab and 3C58 is suggestive of Type IIP, but
a Type Ib/c is also possible.

\subsection{PSR J1119--6127 and G292.2--0.5}

The pulsar PSR J1119--6127 and its surrounding remnant G292.2--0.5 are
notable in that the emission from any PWN is very weak (Crawford et al.
2001; Gonzalez \& Safi-Harb 2003b).
Crawford et al. (2001) argue that the reason for the weak emission is
that the initial spindown time, $\tau$, was small because of the
high magnetic field, so that the internal energy suffered strong
adiabatic losses after the initial period of energy injection.
We concur
with this suggestion.

The radius of the G292.2--0.5 supernova remnant is $10.9d_5$ pc,
where $d_5$ is the distance in units of 5 kpc; the distance is
poorly known and may be in the range $2.5-8$ kpc (Crawford et al. 2001).
The size of the remnant and the relatively small X-ray luminosity,
$(3-4)\times 10^{35}\ergs$ (Pivovaroff et al. 2001), suggest that the
remnant is not a Type IIL/b, again leaving the IIP and Ib/c categories.

\subsection{Discussion}

The basic results on supernova type and pulsar properties are given
in Table 3.
The suggested types are from the 3 main supernova categories and,
in this picture, the main factor in determining the appearance of a young
supernova remnant with a pulsar is the category of the initial supernova.
For the supernova identifications, the circumstellar interaction
plays an important role and can be most clearly seen in the
Type IIL/b and Ib/c events.
For the putative Type IIP events identified here, no circumstellar
interaction has been definitely observed, so there remains some
doubt on the identification.
Another important indicator of supernova type is the composition
of the ejecta, with the presence of slow H distinguishing the IIP from
the IIL/b and Ib/c events.
In some cases, optical emission from the ejecta may be from just a
small fraction of the mass, so it is necessary to assume that the observed
emission is representative of gas at that velocity.

X-ray observations of the supernova remnant shells contain
information on the explosion interactions, but element abundances,
inhomogeneities, and time-dependent effects are important for
the interpretation, which is beyond present considerations.
The X-ray luminosities of the Type IIL/b events are
$L_x\sim 1\times 10^{37}\ergs$, which is higher that the values
for the SNe Ib/c (with $L_x\la 3\times 10^{36}\ergs$), except for
Kes 75 with $L_x\approx 1.8\times 10^{37}\ergs$ (Helfand et al. 2003).
The Type Ib/c events have harder X-ray emission that the IIL/b
events, as expected with their higher mean velocities.

Errors in the ages and values of $P_0$ are difficult to
estimate because of the uncertain pulsar spin evolution
and the detailed properties of the surrounding supernova.
Running a variety of models for individual objects indicates
an uncertainty of order 30\%.

In general, the PWN models show that the observed nebulae are
consistent with a nebula within a factor of a few of 
energy equipartition between particles
and magnetic fields expanding into the inner, freely expanding ejecta
of a supernova.
There is not a clear theoretical reason for energy equipartition, although
Rees \& Gunn (1974) note that there may be a mechanism that keeps the
magnetic field from becoming larger than the equipartition value.

The estimates of 
initial periods are in the range $10-100$ ms,
although there is some concentration around 40 ms.
In cases where the current period is $> 100$ ms, there is
evidence for an initial period that is significantly smaller.
The range of initial periods is comparable to  the values
deduced by van der Swaluw \& Wu (2001) for older remnants in
which the reverse shock wave has compressed the pulsar wind
nebula and the nebula is expanding subsonically.

Table 3 also gives estimates of the current magnetic fields of
the pulsars, based on observations of $P$ and $\dot P$.
It can be seen that neither the magnetic field nor the initial
periods of the pulsars show a clear trend with supernova type.
There are possible reasons for a dependence of the stellar rotation
on the supernova category.
Mass loss from a star is expected to carry away angular momentum, which
could result in slow central rotation.
On the other hand, binary interaction, which might be important for
the IIL/b and Ib/c categories, can increase the angular momentum
of a star.
The results found here do not show evidence for one of these effects
being dominant.

Provided it is not changed by surface nuclear processing, the
atmospheric composition of a young neutron star should reflect
the composition of the fallback material, which can 
vary in the different supernova types.
H is expected in SNe IIP, but not in the types with considerable mass loss.
Slane et al. (2004) find that X-ray observations of the neutron star
in 3C 58 are consistent with a light element atmosphere, as can
occur in a SN IIP by fallback.

\section{SUPERNOVA REMNANTS WITH NON-PULSAR COMPACT OBJECTS}

An important development in recent years has been the discovery
of compact objects that are not normal radio or X-ray pulsars.
These objects have not been detected as radio sources and in
cases where X-ray pulsations are seen (Anomalous X-Ray Pulsars),
the period is relatively long ($\sim 12$ s).
Although the central objects do not generate observable nebulae,
a number of them are surrounded by young supernova remnants,
which can be analyzed for their supernova type.
The young remnants (ages $\sim 10^3$ yr) are listed in Table 4.

\subsection{Cassiopeia A}

The compact object in Cas A belongs to the class of central objects
with thermal X-ray emission, no observable pulsations,
and no surrounding nebula
(Pavlov et al. 2000).
The  case for Cas A being a SN IIL/b was made in Chevalier
\& Oishi (2003).
The positions and expansion rates of the forward and reverse shock
waves can be explained by interaction with a wind.
Outside of the outer shock front is material that can be identified
with clumpy wind gas that was photoionized by the burst of radiation
from the time shock breakout; it extends to a radius of 7 pc.

\subsection{RCW 103}

This remnant was one of the first showing a non-pulsar central
compact object.
The X-ray source shows strong variability, but no pulsations
(Gotthelf et al. 1999).
The distance to the remnant from recent HI absorption line
studies is found to be between 3.1 and 4.6 kpc (Reynoso et al. 2004);
we take a distance of 3.8 kpc.
The radius of the X-ray SNR is 3$^{\prime}$.5, or 5.0 pc.

The remnant radius is in a range where it may still be interacting
with  mass loss from the RSG phase.
Support for this hypothesis comes from an apparent overabundance of
N observed in cooling shock waves at the outer edge of the remnant
(Ruiz 1983).
However, the density of a free wind at 5.0 pc would be
$\rho=4\times 10^{-25}\gcm$ for $D_*=1$,
which is too low to produce the radiative shock waves
and H$_2$ emission that have been observed from the remnant.
A high density shell can be produced if the slow wind has passed
through a termination shock induced by the pressure in the
surrounding medium (\S~\ref{seccsm}).
In this picture, the shell is elongated in the SW--NE direction,
giving the apparent barrel shape of the strong X-ray and radio
emission.
The elongation to the SW, ahead of the shock wave, may show up
as H$_2$ emission in this direction (Oliva et al. 1990).
The radiative shock waves appear where the forward shock wave
encounters the dense parts of the shell; Meaburn \& Allan (1986)
have noted that the velocities of the shocked region suggest
interaction with dense condensations.

The X-ray emission is produced in somewhat lower density parts
of the shell.
The X-ray emission is strongest at the outer parts of the remnant
and the temperature is relatively low.
This is expected for supernova interaction with a shell and is
also observed in SN 1987A (McCray 2003).

An unusual aspect of RCW 103 is the H$_2$ emission from the vicinity
of a young remnant.
Although the emission has been interpreted as interaction with
a molecular cloud (Oliva et al. 1990; Rho et al. 2001), there
is no evidence that the remnant is embedded in a molecular cloud,
and we suggest a circumstellar wind interpretation here.
The presence of the H$_2$ would rule out the possibility that
the progenitor of the supernova was a normal red supergiant,
because of the energetic emission expected at the time of shock breakout.
An alternative is a highly extended RSG, or one with a dense wind
that can sustain a radiation dominated shock wave.
An intriguing possibility is that the shock breakout radiation played
a role in exciting the H$_2$ emission, which cannot be explained by
the excitation from the current radiation field (Rho et al. 2001).

The presence of the extended dense wind in this picture 
places RCW 103 in the SN IIL/b category.

\subsection{Puppis A}

Puppis A has a remarkable morphology, with extended interstellar
interaction, fast O-rich knots, and a central compact X-ray source.
For a distance of 2 kpc, the full remnant diameter is
32 pc; the age derived from the motion of O-rich knots, assumed
to be freely expanding, is 3700 yr (Winkler et al. 1988);
the velocity range of the O-rich knots is $1500-3000\kms$.

An important clue to the supernova type comes from the fact that
there is some H in the O-rich knot, even though the mass ratio
of O/H is $\sim 30$ (Winkler \& Kirshner 1985).
The presence of a small amount of H at a velocity of $1500-3000\kms$
suggests a SN II in which the progenitor has undergone considerable
mass loss, i.e. a SN IIL/b.
The presence of some slow filaments with a N overabundance (Dopita,
Mathewson, \& Ford 1977) is
consistent with material from the RSG wind of the progenitor.
The fact that this gas has been shocked and cooled requires
that it be gas that had passed through the termination shock of
the RSG, or made up of clumps in the RSG wind.
The RSG wind gas is expected to extend out to $\la 7$ pc from the
site of the supernova and it is a prediction of the present model
that slow N-rich material should be present only in the central part
of the remnant.
The large diameter of Puppis A,  32 pc, implies that the outer interaction
is with the interstellar medium.
Blair et al. (1995) have analyzed emission from a shocked cloud
on the eastern side of the remnant; the abundances are deduced to
be close to solar.
It is possible that N-rich ejecta filaments are also present and
can range more widely in the remnant than the RSG wind gas.
Winkler et al. (1989) found a N-rich filament with a velocity
close to $1,000\kms$ in the central part of the remnant.

\subsection{Kes 73}

The young remnant Kes 73 is notable for containing an AXP
(anomalous X-ray pulsar) with a period of 12 s (Vasisht \& Gotthelf 1997),
which is interpreted as a highly magnetized neutron star with
$B\sim 10^{15}$ G.
Sanbonmatsu \& Helfand (1994) estimate a distance to the remnant between
6 and 7.5 kpc from HI absorption.
The remnant radius of $4.7d_7$ pc and swept up mass of
$\sim 8.8d_7^3\Msun$ (Gotthelf \& Vasisht 1997)
are consistent with a SN IIL/b  running into the RSG wind
lost from the progenitor star.
The swept up mass is larger than would be expected for a
SN 1987A-like remnant, although this possibility cannot be 
entirely ruled out.
The models of \S~4 suggest an age of $800-2000$ yr.

\subsection{1E 0102.2--7219 in the SMC}

The remnant E0102 (1E 0102.2--7219) is in the SMC (Small Magellanic
Cloud) at an estimated distance of 59 kpc.
No central compact object or related nebula have been identified
and Gaetz et al. (2000) set an upper limit of $L_x<9\times 10^{33}\ergs$
on the X-ray luminosity of a central PWN.
Using the empirical relation between $L_x$ and $\dot E$ of
Seward \& Wang (1988), we have $\dot E \la 2\times 10^{36}\ergs$.
This is less that the spindown powers of pulsars in other young
remnants (Table 1) and may indicate that the compact object is
of the ``quiet'' variety, although a weak PWN remains a possibility.

The X-ray image of E0102  shows a bright ring
of emission surrounded by a plateau with an outer edge
(Gaetz et al. 2000).
The ring can be identified with the reverse shock wave and the
outer edge with the forward shock;  detailed spectroscopic studies
of the ring emission confirm that it is likely to be a reverse
shock (Flanagan et al. 2004).
The angular diameter of the forward shock is $44^{\prime\prime}$,
corresponding to a radius of 6.3 pc at the distance of the SMC.
The X-ray emission is at the inner boundary for a shell of optical
emission which surrounds the remnant.
This can be best seen in the {\it HST} image of Blair et al. (2000),
which shows that the diameter of the emission is $\sim 60^{\prime\prime}$,
or $R=8.6$ pc.
The ring of X-ray emission can be interpreted in terms of emission
from pure heavy elements (Flanagan et al. 2004) and the remnant
shows optical fast knots of O and Ne, but no H (Blair et al. 2000).

As in the case of Cas A, the evidence is strong that E0102 is
interacting with the dense free wind from the progenitor star and is
the result of a SN IIL/b.
The forward shock wave is at 1.6 times the reverse shock, yet the
outer emission does not show the strong limb brightening that would be
expected if the shock were interacting with a constant density
medium.
The presence of heavy element-rich gas at the reverse shock
is consistent with strong mass loss before the supernova.
The outer ionized region can be attributed to wind material that
was photoionized by the burst of radiation emitted at the time
of shock breakout from a RSG progenitor.
The observation of the He II line in the spectrum
(Tuohy \& Dopita 1983) is indicative
of the hard photoionizing spectrum that is expected from shock
breakout.
The fact that the ionized wind extends out to a radius of 8.6 pc
suggests that the surrounding pressure is relatively low
(see eq. [\ref{csrsg}]), as might occur in the SMC.

   From proper motions at X-ray wavelengths, Hughes, Rakowski, \& Decourchelle
(2000) estimated a free expansion age of 1E0102 of $1000$ yr. From 
the velocities of optical filaments, Eriksen et al. (2001)
estimated an age of $2100$ yr.
In view of the relatively low velocities observed at both optical
and X-ray (Flanagan et al. 2004) wavelengths, we prefer the large
age.
The low velocities also suggest that much of the supernova energy
has been thermalized and it is approaching the blast wave phase.
The comparison of the radius of 1E0102 with interaction models
(Fig. 3) shows that the remnant is close to expectations.   

Table 4 summarizes the properties of the remnants discussed in
this section.
They have all been identified with SN IIL/b, which can be
attributed to the fact that these objects have the strongest
interaction with the surrounding medium and are thus the brightest and
best studied.

\section{SUMMARY AND CONCLUSIONS}

One aim of this paper is to relate the variety of young core
collapse supernova remnants to the various kinds of supernovae.
Four categories of supernovae have been identified, primarily
based on the progenitor mass loss properties.
The Type IIP progenitors are stars with mass $\sim 10-25\Msun$ that explode
with most of their H envelope present.
The mass loss that occurs during the RSG (red supergiant) phase
remains close to the progenitor star; beyond it is a wind bubble
created by the fast wind from the main sequence phase.
During the supernova explosion, the core material is decelerated
by the envelope, and H-rich material is mixed back to a velocity
of a few $100\kms$ or less.

The Type IIL/b events have had extensive mass loss from the
H envelope, so that the envelope is unable to fully decelerate
the core material during the explosion.
The mass loss from the progenitor can extend out $5-7$ pc from
the progenitor.
Observations of a number of these supernovae has shown an overabundance
of N, at least near the star, so this can be taken as an
indication of RSG wind material.
In these supernovae, material moving at several $1000\kms$ has
little or no H present.
Radiation from the time of shock breakout may be able to completely
ionize the surrounding circumstellar medium.
The exact properties of breakout radiation depend on the initial
photospheric radius of the supernova.

The next stage of mass loss is the development of a Wolf-Rayet
star with a fast H-poor wind.
The resulting supernova is of Type Ib/c.
During the Wolf-Rayet phase, the fast wind is typically able to
sweep up the RSG wind, which is left in clumps and at larger radii.
Ionizing radiation is expected at the time of shock breakout, but it
is not capable of fully ionizing the circumstellar material.

The final kind of supernova can be called SN 1987A-like.
SN 1987A had a sufficiently massive H envelope to decelerate the
core and to mix H down to $\la 700\kms$, yet it exploded as a blue
supergiant, which led to an intermediate amount of ionizing radiation
at the time of shock breakout.
The circumstellar medium close to the star is complex, as can be seen in
the current interaction in SN 1987A, but, on a timescale of 100's of years, the
interaction should be with the wind from the RSG phase.
None of the young remnants show properties that point to
a SN 1987A-like event.
This is consistent with the fact that such events do not appear to
contribute substantially to the extragalactic supernova rate, although
they are of low luminosity.
The relatively high metallicity of our Galaxy, compared to the
Large Magellanic Cloud, could also be a factor.

Plausible supernova types for young remnants are given in Table 3.
Two remnants, the Crab and 3C58, are identified with Type IIP.
Both show slow moving H rich gas and appear to have low density surroundings.
However, in neither case has the interaction with the surroundings been
observed, which introduces some doubt into the identification.
G292.0+1.8 and G11.2-0.3 show evidence for strong circumstellar interaction
on a scale of 5--7 pc, which is the region expected to be occupied by
the RSG wind in the case of a SN IIL/b.
For G292.0+1.8, the lack of H in ejecta moving at $\sim 2000\kms$ is further
evidence for this supernova identification.

The  nebulae 0540--69, Kes 75, and MSH 15-52 are plausibly
identified with Type Ib/c remnants.
They all have expanded with an average velocity $\sim 10,000\kms$ to radii
$>10$ pc, which is inconsistent with interaction with the dense wind
expected around a Type IIL/b.
In the cases of 0540--69 and MSH 15-52, there is evidence for N-rich
clumps in the outer parts of the remnants which can be identified with
RSG material that has been accelerated outward by interaction with the
Wolf-Rayet progenitor wind.
An expectation for this scenario is that the ejecta are lacking H, for
which there is some evidence in the case of 0540--69.

Models for the PWNe are developed  based on the assumption of interaction
with normal supernova ejecta and the use of the minimum energy from synchrotron
emission to set a constraint on the internal energy.
Successful models for the observed nebulae have an internal energy
which is within a factor of a few of the minimum energy.
The models  provide an estimate of age of the remnant.
In the case of 3C58, the age is $2400\pm 500$ yr, inconsistent with the
identification of 3C58 as the remnant of SN 1181.
The larger age is indicated by the internal energy in the PWN, the
expansion of the PWN,  the mass of swept-up thermal gas, and the
temperature of shocked ejecta, and is
consistent with the relatively low expansion observed at optical and
radio wavelengths.
Models for the PWNe in MSH 15-52 and G11.2-0.3 are consistent with
their identification the SN 185 and SN 386, respectively.
The age of MSH 15-52 is better specified by the models than that of G11.2-0.3.

The models allow estimates of the initial rotation periods of the
pulsars, which are found to be in the range $10-100$ ms (Table 3).
The pulsars with current periods $>100$ ms show evidence for substantial
spindown in order to be consistent with the PWN properties.
The initial periods show no strong trend with supernova type.
Table 3 also shows that the magnetic fields of the pulsars
are not related to the supernova type.
The presence of pulsars in the remnants of the various supernova
types and the similar properties of the central pulsars is
an indication that the supernova types are not a mass sequence,
as would be expected in a single star origin for the supernovae.
In a binary origin for a substantial fraction of Type IIL/b
and Type Ib/c supernovae, they overlap with the masses of
Type IIP events so that the core masses can be in a similar range.

The lack of a relation between pulsar properties and supernova type
is also an indication that mass fallback is not an important
process for the basic neutron star properties.
The estimated initial rotation periods of the pulsars limits
the amount of fallback that can occur at near the Keplerian
velocity.
However, even a minute amount of fallback is important for
the composition of the neutron star atmosphere.
A H rich atmosphere is commonly assumed for young neutron stars.
In the Type IIL/b and Ib/c events, material falling on
the neutron is likely to be lacking H.

A number of young remnants without normal pulsars as central
objects appear to fall into the Type IIL/b category.
However, the lack of Type IIP and Ib/c events may be simply
due to the fact that these remnants have weak circumstellar
interaction and are faint.

The PWNe discussed in the here have ages $\sim 10^3$ yr.
The finding of PWNe in the light from supernovae with ages $\sim 10$ yr
would directly solve the problem of identifying a PWN with a particular
supernova type and would substantially extend the known evolutionary
sequence of PWNe.
In this context, the recent discovery of  compact, central radio
emission in the Type IIn SN 1986J (Bietenholz, Bartel, \& Rupen 2004)
is of special interest.
The source shows an inverted spectrum with a turnover near 20 GHz,
as expected for free-free absorption
by material in the surrounding supernova.
The amount of absorption to a central PWN depends on the density
structure of the supernova (\S~2.1), and also on the physical
conditions in the gas.
Sensitive observations of supernovae at late times provide the
opportunity to study very young PWNe.

\acknowledgments
I am grateful to Richard Mellon for contributing to the
thin shell models for pulsar wind nebulae, to the referee, David Branch,
for helpful comments on the manuscript, and to Parviz Ghavamian
and David Helfand for useful discussions.
Support for this work was provided in part by NASA grant NAG5-13272
and NSF grant AST-0307366.
I am also grateful for the stimulating atmosphere and support provided
by the ISSI (International Space Science Institute, Bern) workshop on the
``Physics of Supernova Remnants in the XMM-Newton, Chandra and INTEGRAL Era.''

\clearpage

\vspace*{1 cm}
\noindent{Table 1. Properties of Pulsars and their Nebulae }

\begin{tabular}{ccccccccc}
\hline
PSR & Supernova &  Distance  & $\dot E/I_{45}$ & $P$ & $P/2\dot P$   & $R_{PWN}$ & $R_{SNR}$  \\
     &    Remnant  & (kpc)  & (ergs s$^{-1}$)& (ms) & (yr) &  (pc)  &  (pc)  &      \\
\hline
B0531+21 &  Crab &  2 &  $4.7\times 10^{38}$   &   33   & 1240  & 2 & --   \\
J0205+64 &  3C 58   &3.2  &  $2.7\times 10^{37}$  &  66 & 5390  & 3.3  & --    \\
B0540--69 &  N158A &  50 &  $1.5\times 10^{38}$   & 50 & 1660  &  0.9 & 9   \\
J1846--03  &  Kes 75 &19 &  $8.3\times 10^{36}$   & 325&  723 & 1.4   & 9.7  \\
B1509--58 &  MSH 15-52 & 5.2 &  $1.8\times 10^{37}$   &  150  & 1700 & 5.5   & 19   \\
J1124--59 &  G292.0+1.8  & 6  &  $1.2\times 10^{37}$  &   135  & 2890 & 3.5  & 7  \\
J1811--19 &  G11.2--0.3 & 5  &  $6.4\times 10^{36}$   &  65  & 24,000  & 0.9 & 3.3  \\
J1930+19  &  G54.1+0.3  &  $\sim 5$  &$1.2\times 10^{37}$  & 137 & 2890 &  1.2 & --   \\
J1119--61  &  G292.2--0.5  &  $\sim 6$  &$2.3\times 10^{36}$  & 408 & 1606 &  -- & 11   \\
\hline

\end{tabular}

\vspace*{1 cm}
\noindent{Table 2. Minimum internal energies of pulsar wind nebulae }

\begin{tabular}{ccccccc}
\hline
 Supernova &  $p_1$  & $p_2$ & $\nu_b$ & $L_{\nu b}$   & $E_{min}$  &  Reference\\
   Remnant  &       &        & (Hz)  & (ergs s$^{-1}$ Hz$^{-1}$) &  (ergs)  & \\
\hline
 Crab &  1.6 & 3.0  &  $1\times 10^{13}$  & $9.5\times 10^{23}$  &$6\times 10^{48}$ & 1\\
 3C 58   & 1.2 & 3.0  &  $5\times 10^{10}$  & $3.0\times 10^{23}$  &$1\times 10^{48}$ &2 \\
0540--69  & 1.5 & 2.6  &  $2\times 10^{13}$  & $3.0\times 10^{22}$  &$5\times 10^{47}$ &1 \\
 Kes 75 &  1.0 & 3.0  &  $1\times 10^{13}$  & $1.2\times 10^{23}$  &$1\times 10^{48}$ &3 \\
 MSH 15-52 & 1.4 & 3.1  &  $3\times 10^{13}$  & $4.4\times 10^{21}$  &$1.5\times 10^{48}$ &4 \\
 G292.0+1.8  & 1.1 & 2.8  &  $3\times 10^{10}$  & $2.0\times 10^{23}$  &$1\times 10^{48}$&5 \\
 G11.2--0.3 & 1.5 & 2.46  &  $8\times 10^{9}$  & $7.8\times 10^{21}$  &$3\times 10^{46}$&6 \\
 G54.1+0.3  & 1.26 & 2.8  &  $5\times 10^{11}$  & $6.7\times 10^{21}$  &$8\times 10^{46}$ & 7 \\
\hline

\end{tabular}  

\noindent References:
(1) Manchester et al. 1993;
(2) Green \& Scheuer 1992;
(3) Blanton \& Helfand 1996;
(4) Gaensler et al. 2002;
(5) Gaensler \& Wallace 2003;
(6) Roberts et al. 2003;
(7) Lu et al. 2002.

\clearpage
\vspace*{1.7 cm}
\noindent{Table 3.   Properties deduced from models}

\begin{tabular}{cccccccc}
\hline
Supernova &   Supernova &  Age    &  $P_0$  & $B$  \\
Remnant   &    Type &   (year) &  (ms)  & ($10^{12}$ G)   \\
\hline
Crab &  IIP &   950    & 20  &  4 \\
3C 58 &  IIP   &   2400    & 50  &  4 \\
0540--69 &  Ib/c &   800   & 40  &  5  \\
Kes 75 &  Ib/c &   1000  & 30   &  48  \\
MSH 15--52 &  Ib/c &   1700    & 10  &  14  \\
G292.0+1.8 &  IIL/b &   3200    & 40 &  10 \\
G11.2--0.3 & IIL/b  &   1600  & 60  & 2  \\
G54.1+0.3 &  IIP,Ib/c &   1500    &  100  &  10  \\
G292.2--0.5 &  IIP,Ib/c &   1700    &  $\ll 200$  &  41  \\
\hline
\end{tabular}

\vspace*{1.7 cm}
\noindent{Table 4.   Properties of remnants without normal pulsars}

\begin{tabular}{cccccccc}
\hline
Supernova &  Distance  &  Supernova &  Age    &  Radius    \\
Remnant   & (kpc)   &   Type &   (year) &  (pc)     \\
\hline
Cas A &   3.4    &   IIL/b  &   320    & 2.5   \\
RCW 103 &   3.8   &    IIL/b  &   --    & 5.0   \\
Puppis A &   2   &  IIL/b  &   3700    & 16   \\
Kes 73  &   7   &  IIL/b   &  --   &     4.7    \\
1E 0102 &    59  &  IIL/b  &   1000--2100    & 6.3   \\
\hline
\end{tabular}

\clearpage

\begin{figure}[!hbtp]
\plotone{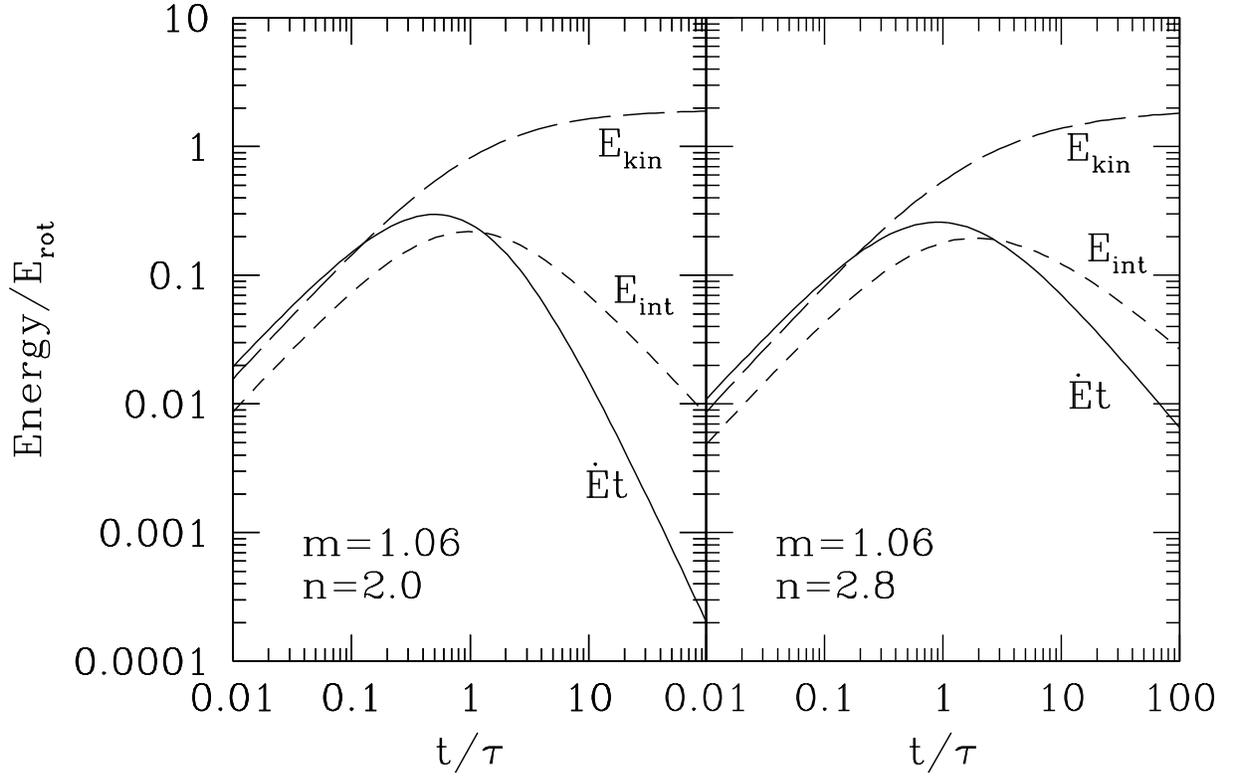}
\figcaption{The evolution of pulsar nebula internal energy, $E_{int}$,
kinetic energy of the swept-up shell, $E_{kin}$, and the current
$\dot E t$ divided by the initial rotational energy, $E_{rot}$.
The 2 models are characterized by the power law index of the
supernova density profile, $m$, and the pulsar braking index, $n$.
The reference time $\tau$ is the initial spindown time of the pulsar.}
\end{figure}

\clearpage

\begin{figure}[!hbtp]
\plotone{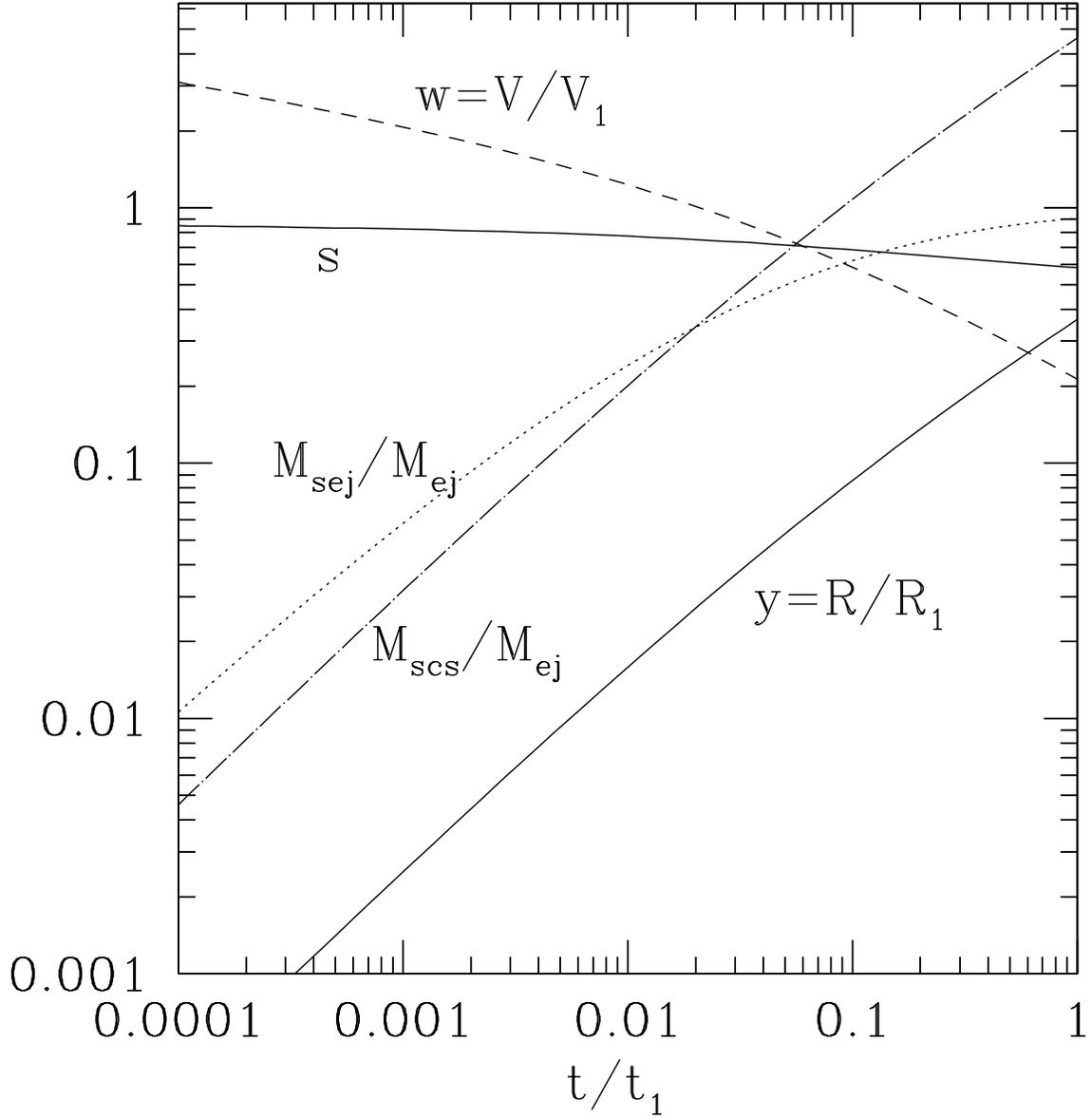}
\figcaption{Results for the thin shell circumstellar 
interaction model described in \S~4.
The dimensionless variables $y$ (radius) and $w$ 
(velocity) are given as a function
of time; $s$ is the deceleration parameter for the shell.
The swept up circumstellar mass, $M_{scs}$, and ejecta mass,
$M_{sej}$, are given relative to the total ejecta mass, $M_{ej}$.}
\end{figure}

\clearpage

\begin{figure}[!hbtp]
\plotone{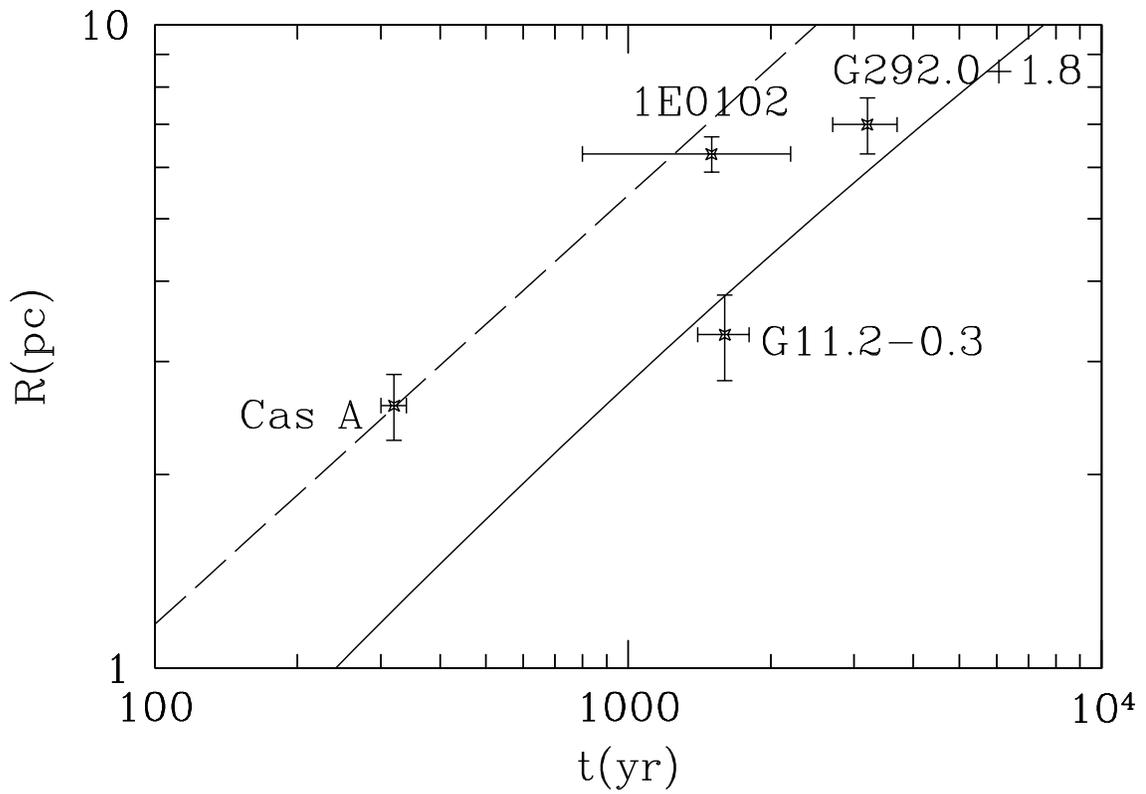}
\figcaption{Models for thin shell expansion (solid line) and blast
wave expansion (dashed line) in a circumstellar wind, using typical
parameters (see \S~4).
The observational points are for young remnants that may be 
expanding in a wind.}
\end{figure}


\clearpage


\begin{thebibliography} {}


\bibitem[]{2176}
Bandiera, R., Pacini, F., \& Salvati, M. 1983, A\&A, 126, 7

\bibitem[Basko(1994)]{1994ApJ...425..264B} Basko, M.\ 1994, \apj, 425, 264


\bibitem[]{2182}
Bietenholz, M. F., Bartel, N., \& Rupen, M. P. 2004, Science, 304, 1947


\bibitem[]{2186}
Bietenholz, M. F., Kassim, N. E., \& Weiler, K. W. 2001, ApJ, 560, 772

\bibitem[Blair, Raymond, Long, \& Kriss(1995)]{1995ApJ...454L..35B} Blair,
W.~P., Raymond, J.~C., Long, K.~S., \& Kriss, G.~A.\ 1995, \apjl, 454, L35


\bibitem[Blair et al.(2000)]{2000ApJ...537..667B} Blair, W.~P.~et al.\
2000, \apj, 537, 667   
\bibitem[]{2195}
Blanton, E. L., \& Helfand, D. J. 1996, ApJ, 470, 961

\bibitem[Blinnikov et al.(2003)]{2003fthp.conf...23B} Blinnikov, S.,
Chugai, N., Lundqvist, P., Nadyozhin, D., Woosley, S., \& Sorokina, E.\
2003, in From
Twilight to Highlight: The Physics of Supernovae, Eds. W. Hillebrandt
\& B. Leibundgut, (Berlin: Springer),  23
\bibitem[]{2203}
Bocchino, F., Warwick, R. S., Marty, P., Lumb, D., Becker, W., \& Pigot, C.
2001, A\&A, 369, 1078

\bibitem[]{2207}
Bucciantini, N., Blondin, J. M., Del Zanna, L., \& Amato, E. 2003, A\&A, 405, 617
\bibitem[]{2209}
Camilo, F., et al. 2000, ApJ, 541, 367

\bibitem[Camilo et al.(2002a)]{C02a}
Camilo, F., Lorimer, D. R., Bhat, N. D. R., Gotthelf, E. V.,
Halpern, J. P., Wang, Q. D., Lu, F. J., \& Mirabal, N. 2002a, ApJ, 574, L71
\bibitem[Camilo et al.(2002b)]{Cet02b}
Camilo, F., Manchester, R. N., Gaensler, B. M., Lorimer, D. R.,
\& Sarkissian, J. 2002b, ApJ, 567, L71
\bibitem[]{2218}
Cappellaro, E., Barbon, R., \&
Turatto, M.  2004, in IAU Colloquium 192, Supernovae: 10 Years of
SN 1993J, eds. J.M. Marcaide \& K.W. Weiler (Berlin: Springer), in press
(astro-ph/0310859)
\bibitem[]{2223} Cappellaro, E.,
Turatto, M., Benetti, S., Tsvetkov, D.~Y., Bartunov, O.~S., \& Makarova,
I.~N.\ 1993, \aap, 273, 383
\bibitem[]{2226}
Cappellaro, E.,
Turatto, M., Tsvetkov, D. Y., Bartunov, O. S., Pollas, C., Evans, R., \&
Hamuy, M.  1997, \aap, 322, 431
\bibitem[Chevalier(1977)]{C77}
Chevalier, R. A. 1977, in Supernovae, ed. D. N. Schramm (Dordrecht: Reidel), 53

\bibitem[]{2233}
Chevalier, R. A. 1982, ApJ, 259, 302
\bibitem[Chevalier(1989)]{1989ApJ...346..847C} Chevalier, R.~A.\ 1989,
\apj, 346, 847
\bibitem[]{2237} Chevalier, R.~A.\ 2003, in From
Twilight to Highlight: The Physics of Supernovae, Eds. W. Hillebrandt
\& B. Leibundgut, (Berlin: Springer), 299
\bibitem[]{2240}
Chevalier, R. A. 2004, AdSpR, 33, 456
\bibitem[]{2242} Chevalier,
R.~A.~\& Emmering, R.~T.\ 1989, \apjl, 342, L75
\bibitem[]{2244}
Chevalier, R. A.,  \& Fransson, C. 1992, ApJ, 395, 540
\bibitem[]{2246}
Chevalier, R. A.,  \& Imamura, J. N. 1983, ApJ, 270, 554

\bibitem[]{2249}
Chevalier, R. A.,  \& Oishi, J. N. 2003, ApJ, 593, L23


\bibitem[]{2253}
Clark, D. H., \& Stephenson, F. R. 1977, The Historical Supernovae
(New York: Pergamon)
\bibitem[Colgate(1971)]{1971ApJ...163..221C} Colgate, S.~A.\ 1971, \apj,
163, 221
\bibitem[]{2258}
Crawford, F., et al. 2001, ApJ, 554, 152
\bibitem[]{}
Crotts, A., Sugerman, B., Lawrence, S., \& Kunkel, W. 2001, in Young
Supernova Remnants, ed. S. S. Holt \& U. Hwang (Melville: AIP),  137
\bibitem[Dopita, Mathewson, \& Ford(1977)]{1977ApJ...214..179D} Dopita,
M.~A., Mathewson, D.~S., \& Ford, V.~L.\ 1977, \apj, 214, 179

\bibitem[]{2264} Dopita, M.~A.~\&
Tuohy, I.~R.\ 1984, \apj, 282, 135
\bibitem[]{2266}
Dwarkadas, V. 2001, JKAS, 34, 243
\bibitem[Elmhamdi et al.(2003)]{2003MNRAS.338..939E} Elmhamdi, A., et al.\
2003, \mnras, 338, 939

\bibitem[]{2271}
Eriksen, K. A., Morse, J. A., Kirshner, R. P., \& Winkler, P. F. 2001,
in Young Supernova Remnants, ed. S. S. Holt \& U. Hwang (Melville: AIP), 193

\bibitem[]{2275}
Fesen, R. A., Kirshner, R. P., \& Becker, R. H. 1988, in
Supernova Remnants and the Interstellar Medium,
ed. R. S. Roger, \& T. L. Landecker (Cambridge: Cambridge University Press), 55

\bibitem[]{2280}
Fesen, R. A., Shull, M. J., \& Hurford, A. P. 1997, AJ, 113, 354

\bibitem[]{2283}
Flanagan, K. A. , Canizares, C. R.,  Dewey, D.,  Houck, J. C.,
Fredericks, A. C.,  Schattenburg, M. L.,  Markert, T. H., \& Davis, D. S.
2004, ApJ, 605, 230

\bibitem[]{2288}
Frail, D. A., \& Moffett, D. A. 1993, ApJ, 408, 637
\bibitem[Fransson \& Chevalier(1989)]{1989ApJ...343..323F} Fransson, C.,~\&
Chevalier, R.~A.\ 1989, \apj, 343, 323
\bibitem[]{2292}
Fransson, C., et al. 2004, ApJ, in press (astro-ph/0409439)
\bibitem[]{2294}
Gaensler, B. M.,  \& Wallace, B. J. 2003, ApJ, 594, 326
\bibitem[Gaensler et al.(2001)]{G01}
Gaensler, B. M., Arons, J., Kaspi, V. M., Pivovaroff, M. J.,
Kawai, N., \& Tamura, K. 2002, ApJ, 569, 878
\bibitem[]{2299}
Gaensler, B. M., Brazier, K. T. S., Manchester, R. N., Johnston, S., \&
Green, A. J. 1999, MNRAS, 305, 724

\bibitem[Gaetz et al.(2000)]{2000ApJ...534L..47G} Gaetz, T.~J., Butt,
Y.~M., Edgar, R.~J., Eriksen, K.~A., Plucinsky, P.~P., Schlegel, E.~M., \&
Smith, R.~K.\ 2000, \apjl, 534, L47

\bibitem[Garcia-Segura, Langer, \& Mac Low(1996)]{1996A&A...316..133G}
Garcia-Segura, G., Langer, N., \& Mac Low, M.-M.\ 1996, \aap, 316, 133
\bibitem[]{2309}
Ghavamian, P., Hughes, J. P., \& Williams, T. B. 2004, ApJ, submitted
\bibitem[]{2311}
Gonzalez, M., \& Safi-Harb, S. 2003a, ApJ, 583, L91
\bibitem[]{2313}
Gonzalez, M., \& Safi-Harb, S. 2003b, ApJ, 591, L143
\bibitem[Gotthelf, Petre, \& Vasisht(1999)]{1999ApJ...514L.107G} Gotthelf,
E.~V., Petre, R., \& Vasisht, G.\ 1999, \apjl, 514, L107

\bibitem[]{2318} Gotthelf, E.~V., \& Vasisht, G.
1997, ApJ, 486, L133
\bibitem[Gotthelf, Vasisht, Boylan-Kolchin, \&
Torii(2000)]{2000ApJ...542L..37G} Gotthelf, E.~V., Vasisht, G.,
Boylan-Kolchin, M., \& Torii, K.\ 2000, \apjl, 542, L37


\bibitem[]{2325}
Green, D. A. 2004, A Catalogue of Galactic Supernova Remnants,
Mullard Radio Astronomy Observatory, Cavendish Laboratory, Cambridge,
UK (available at ``http://www.mrao.cam.ac.uk/surveys/snrs/'')

\bibitem[]{2330}
Green, D. A., \& Scheuer, P. A. G. 1992, MNRAS, 258, 833
\bibitem[Green, et al.(1988)]{G88}
Green, D. A., Gull, S. F., Tan, S. M., \& Simon, A. J. B. 1988,
MNRAS, 231, 735
\bibitem[]{2335} Hamuy, M.\ 2003, \apj, 582, 905
\bibitem[]{2336} Heger, A., Fryer, C.~L.,
Woosley, S.~E., Langer, N., \& Hartmann, D.~H.\ 2003, \apj, 591, 288



\bibitem[Helfand et al.(2003)]{H03}
Helfand, D. J., Collins, B. F., \& Gotthelf, E. V. 2003, ApJ, 572, 783

\bibitem[]{2346} Herant, M.,~\&
Woosley, S.~E.\ 1994, \apj, 425, 814
\bibitem[Houck \& Fransson(1996)]{1996ApJ...456..811H} Houck, J.~C.,~\&
Fransson, C.\ 1996, \apj, 456, 811

\bibitem[]{2351}
Hughes, J. P., Rakowski, C. E.,  \&
Decourchelle, A. 2000, ApJ, 543, L61
\bibitem[]{2354}
Hughes, J. P., Slane, P. O., Park, S., Roming, P. W. A., \& Burrows, D. N.
2003, ApJ, 591, L139

\bibitem[Hughes et al.(2001)]{H01}
Hughes, J. P., et al. 2001, ApJ, 559, L153

\bibitem[]{2361}
Hwang, U., Petre, R., Holt, S.~S., \& Szymkowiak, A.~E.\ 2001, \apj, 560,
742    


\bibitem[]{2366} Iwamoto, K., Nomoto,
K., Hoflich, P., Yamaoka, H., Kumagai, S., \& Shigeyama, T.\ 1994, \apjl,
437, L115
\bibitem[Iwamoto et al.(1997)]{1997ApJ...477..865I} Iwamoto, K., Young,
T.~R., Nakasato, N., Shigeyama, T., Nomoto, K., Hachisu, I., \& Saio, H.\
1997, \apj, 477, 865
\bibitem[]{2372}
Jun, J.-I. 1998, ApJ, 499, 282

\bibitem[]{2375}
Kaspi, V. M., Manchester, R. N., Siegman, B., Johnston, S., \& Lyne, A. G.
1994, ApJ, 422, L83

\bibitem[]{2379}
Kennel, C. F., \& Coroniti, F. V. 1984, ApJ, 283, 694

\bibitem[]{2382} Kifonidis, K., Plewa, T., Janka, H.-T.,
\& M{\" u}ller, E.\ 2003, \aap, 408, 621


\bibitem[Kirshner et al.(1989)]{K89}
Kirshner, R. P., Morse, J. A., Winkler, P. F., \& Blair, W. P. 1989,
ApJ, 342, 260
\bibitem[Klein \& Chevalier(1978)]{1978ApJ...223L.109K} Klein, R.~I.,~\&
Chevalier, R.~A.\ 1978, \apjl, 223, L109
\bibitem[]{2391} Kozma, C.,~\&
Fransson, C.\ 1998, \apj, 497, 431
\bibitem[Lu et al.(2002)]{L02}
Lu, F. J., Wang, Q. D., Aschenbach, B., Durouchoux, P.,
\& Song, L. M. 2002, ApJ, 568, L49

\bibitem[]{2397} Lundqvist, P.,~\&
Fransson, C.\ 1996, \apj, 464, 924

\bibitem[]{2400}
Lyne, A. G., Pritchard, R. S., Graham-Smith, F., \& Camilo, F. 1996,
Nature, 381, 497

\bibitem[]{2404}
Lyne, A. G., Pritchard, R. S., \& Smith, F. G. 1988, MNRAS, 233, 667
\bibitem[Manchester, Staveley-Smith, \&
Kesteven(1993)]{1993ApJ...411..756M} Manchester, R.~N., Staveley-Smith, L.,
\& Kesteven, M.~J.\ 1993, \apj, 411, 756
\bibitem[MacFadyen, Woosley, \& Heger(2001)]{2001ApJ...550..410M}
MacFadyen, A.~I., Woosley, S.~E., \& Heger, A.\ 2001, \apj, 550, 410
\bibitem[]{2411}
Manchester, R. N., Staveley-Smith, L., \& Kesteven, M. J. 1993, ApJ, 411, 756
\bibitem[]{2413}
Mathewson, D. S., Dopita, M. A., Tuohy, I. R., \& Ford, V. L.
1980, ApJ, 242, L73
\bibitem[Matzner \& McKee(1999)]{MM99}
Matzner, C. D., \& McKee, C. F. 1999, ApJ, 510, 379
\bibitem[]{2418} Maund, J.~R., Smartt,
S.~J., Kudritzki, R.~P., Podsiadlowski, P., \& Gilmore, G.~F.\ 2004, \nat,
427, 129
\bibitem[McCray(2003)]{2003sgrb.conf..219M} McCray, R.\ 2003, in LNP Vol.~598:
Supernovae and Gamma-Ray Bursters, ed. K.W. Weiler (Berlin: Springer), 219
\bibitem[Meaburn \& Allan(1986)]{1986MNRAS.222..593M} Meaburn, J.,~\& Allan,
P.~M.\ 1986, \mnras, 222, 593
\bibitem[Murdin \& Clark(1979)]{MC79}
Murdin, P., \& Clark, D. H. 1979, MNRAS, 189, 501
\bibitem[]{2427}
Murray, S. S., Slane, P. O., Seward, F. D., Ransom, S. M., \& Gaensler, B. M.
2002, ApJ, 568, 226
\bibitem[]{2430}
Nomoto, K., Iwamoto, K., \& Suzuki, T. 1995, Phys. Rep., 256, 173
\bibitem[]{2432}
Nomoto, K., et al. 2001, in Supernovae and Gamma-Ray Bursts,
eds. M. Livio, N. Panagia, \& K. Sahu (Cambridge: CUP), p. 144
\bibitem[]{2435} Nomoto, K., Sugimoto,
D., Sparks, W.~M., Fesen, R.~A., Gull, T.~R., \& Miyaji, S.\ 1982, \nat,
299, 803
\bibitem[Oliva, Moorwood, \& Danziger(1990)]{1990A&A...240..453O} Oliva,
E., Moorwood, A.~F.~M., \& Danziger, I.~J.\ 1990, \aap, 240, 453

\bibitem[Ostriker \& Gunn(1971)]{OG71}
Ostriker, J. P., \& Gunn, J. E. 1971, ApJ, 164, L95

\bibitem[]{2444}
Pacholczyk, A. G. 1970, Radio Astrophysics (San Francisco: Freeman)

\bibitem[]{2447}
Park, S., et al. 2002, ApJ, 564, L39
\bibitem[]{2449} Pastorello, A.~et
al.\ 2004, \mnras, 347, 74

\bibitem[]{2452}
Pavlov, G. G., Zavlin, V. E., Aschenbach, B., Tr\"umper, J., \&
Sanwal, D.\ 2000,  \apj, 531, L53

\bibitem[]{2456}
Pooley, D. et al.\ 2002,  \apj, 572, 932
\bibitem[]{2458}
Reed, J. E., Hester, J. J., Fabian, A. C., \& Winkler, P. F. 1995, ApJ, 440, 706
\bibitem[]{2460}
Rees, M. J., \& Gunn, J. E. 1974, MNRAS, 167, 1

\bibitem[]{2463}
Reynolds, S. P. 1985, ApJ, 291, 152   

\bibitem[]{2466}
Reynolds, S. P., \& Chevalier, R. A. 1984, ApJ, 278, 630

\bibitem[Reynoso et al.(2004)]{2004PASA...21...82R} Reynoso, E.~M., Green,
A.~J., Johnston, S., Goss, W.~M., Dubner, G.~M., \& Giacani, E.~B.\ 2004,
PASA, 21, 82
\bibitem[Rho, Reach, Koo, \& Cambresy(2001)]{2001AIPC..565..197R} Rho, J.,
Reach, W.~T., Koo, B.-C., \& Cambresy, L.\ 2001, in Young
Supernova Remnants, ed. S. S. Holt \& U. Hwang (Melville: AIP),  197

\bibitem[Roberts et al.(2003)]{R03}
Roberts, M. S. E., Tam, C. R., Kaspi, V. M., Lyutikov, M., Vasisht, G.,
Pivovaroff, M., Gotthelf, E. V., \& Kawai, N. 2003, ApJ, 588, 992   

\bibitem[Ruiz(1983)]{1983AJ.....88.1210R} Ruiz, M.~T.\ 1983, \aj, 88, 1210

\bibitem[]{2482}
Sanbonmatsu, K. Y., \& Helfand, D. J. 1992, AJ, 104, 2189

\bibitem[Schaller, Schaerer, Meynet, \& Maeder(1992)]{1992A&AS...96..269S}
Schaller, G., Schaerer, D., Meynet, G., \& Maeder, A.\ 1992, \aaps, 96, 269
\bibitem[]{2487}
Seward, F. D., Harnden, F. R., Jr., Murdin, R., \& Clark, D. H.
1983, ApJ, 267, 698
\bibitem[]{2490}
Seward, F. D.,  \& Wang, Z.-R.
1988, ApJ, 332, 199
\bibitem[]{2493} Shigeyama, T.~\&
Nomoto, K.\ 1990, \apj, 360, 242
\bibitem[]{2495} Shigeyama, T.,
Iwamoto, K., Hachisu, I., Nomoto, K., \& Saio, H.\ 1996, in IAU Colloq.~145:
Supernovae and Supernova Remnants, eds. R. McCray \& Z. Wang
(Cambridge: CUP), 129
\bibitem[]{2499}
Slane, P., Helfand, D. J., van der Swaluw, E., \& Murray, S. S.
2004, ApJ, in press (astro-ph/0405380)

\bibitem[]{2503} Smartt, S.~J., Maund,
J.~R., Gilmore, G.~F., Tout, C.~A., Kilkenny, D., \& Benetti, S.\ 2003,
\mnras, 343, 735
\bibitem[Stephenson \& Green(2002)]{SG02}
Stephenson, F. R., \& Green, D. A. 2002, Historical Supernovae and
their Remnants (Oxford Univ. Press: Oxford)
\bibitem[]{2509}
Tam, C., \& Roberts, M. S. E. 2003, ApJ, 598, L27  
\bibitem[Tan et al.(2002)]{T02}
Tam, C., Roberts, M. S. E., \& Kaspi, V. M. 2002, ApJ, 572, 202  

\bibitem[Torii, Tsunemi, Dotani, \& Mitsuda(1997)]{1997ApJ...489L.145T}
Torii, K., Tsunemi, H., Dotani, T., \& Mitsuda, K.\ 1997, \apjl, 489, L145

\bibitem[]{2517}
Torii, K., et al. 1999, \apjl, 523, L69

\bibitem[Tuohy \& Dopita(1983)]{1983ApJ...268L..11T} Tuohy, I.~R.,~\&
Dopita, M.~A.\ 1983, \apjl, 268, L11

\bibitem[]{2523} van der Swaluw, E., \& Wu, Y. 2001, ApJ, 555, L49


\bibitem[]{swal00}
van der Swaluw, E., Achterberg, A., Gallant, Y. A., \& T\'oth, G. 2001,
A\&A, 380, 309

\bibitem[]{2530} Van Dyk, S.~D.,
Garnavich, P.~M., Filippenko, A.~V., H{\" o}flich, P., Kirshner, R.~P.,
Kurucz, R.~L., \& Challis, P.\ 2002, \pasp, 114, 1322

\bibitem[]{2534} Van Dyk,
S.~D., Li, W., \& Filippenko, A.~V.\ 2003, \pasp, 115, 1289
\bibitem[]{2536} Vasisht, G., \& Gotthelf, E.~V.
1997, ApJ, 486, L129

\bibitem[]{2539}
Wellstein, S., \& Langer, N. 1999, A\&A, 350, 148

\bibitem[Winkler \& Kirshner(1985)]{1985ApJ...299..981W} Winkler, P.~F.,~\&
Kirshner, R.~P.\ 1985, \apj, 299, 981

\bibitem[Winkler, Kirshner, Hughes, \&
Heathcote(1989)]{1989Natur.337...48W} Winkler, P.~F., Kirshner, R.~P.,
Hughes, J.~P., \& Heathcote, S.~R.\ 1989, \nat, 337, 48


\bibitem[Winkler, Tuttle, Kirshner, \& Irwin(1988)]{1988srim.conf...65W}
Winkler, P.~F., Tuttle, J.~H., Kirshner, R.~P., \& Irwin, M.~J.\ 1988, IAU
Colloq.~101: Supernova Remnants and the Interstellar Medium,
ed. R. S. Roger \& T. L. Landecker (Cambridge: CUP), 65

\bibitem[]{2555}
Woosley, S. E. 1988, ApJ, 330, 218

\bibitem[]{2558}
Zhang, W., Marshall, F. E., Gotthelf, E. V., Middleditch, J., \& Wang, Q. D.
2001, ApJ, 554, L177

%


\end{thebibliography}
\end{document}